\documentclass[fleqn,usenatbib]{mnras}

\usepackage{newtxtext,newtxmath}
\usepackage{xcolor}

\usepackage[T1]{fontenc}
\usepackage{comment}
\usepackage{subcaption}
\DeclareRobustCommand{\VAN}[3]{#2}
\let\VANthebibliography\thebibliography
\def\thebibliography{\DeclareRobustCommand{\VAN}[3]{##3}\VANthebibliography}

\usepackage{graphicx}	
\usepackage{amsmath}	
\usepackage{comment}

\title[High-z JWST Spectra]{Adding Value to JWST Spectra and Photometry: Stellar Population and Star Formation Properties of Spectroscopically Confirmed JADES and CEERS Galaxies at $z > 7$}

\author[Qiao Duan et al.]{
Qiao Duan,$^{1}$
Christopher J. Conselice,$^{1}$\thanks{E-mail: conselice@manchester.ac.uk}
Qiong Li,$^{1}$
Thomas Harvey,$^{1}$ 
Duncan Austin,$^{1}$
Katherine Ormerod,$^{1}$  \and
James Trussler,$^{1}$
Nathan Adams$^{1}$
\\
$^{1}$ Jodrell Bank Centre for Astrophysics, University of Manchester, Oxford Road, Manchester UK}

\date{Accepted XXX. Received YYY; in original form ZZZ}

\pubyear{2023}

\begin{document}
\label{firstpage}
\pagerange{\pageref{firstpage}--\pageref{lastpage}}
\maketitle

\begin{abstract}

    In this paper, we discuss measurements of the stellar population and star forming properties for 43 spectroscopically confirmed publicly available high-redshift \( z > 7 \) JWST galaxies in the JADES and CEERS observational programs. 
We carry out a thorough study investigating the relationship between spectroscopic features and photometrically derived ones, including from spectral energy distribution (SED) fitting of models, as well as morphological and structural properties. 
We find that the star formation rates (SFRs) measured from H$\beta$ line emission are higher than those estimated from Bayesian SED fitting and UV luminosity, with ratios \( \text{SFR}_{\text{H}\beta} / \text{SFR}_{\text{UV}} \)  ranging from $\sim 2 - 13$.
This is a sign that the star formation history is consistently rising given the timescales of H$\beta$ vs UV star formation probes. In addition, we investigate how well equivalent widths (EWs) of H\(\beta\) \(\lambda4861\), [O~\textsc{iii}] \(\lambda4959\), and [O~\textsc{iii}] \(\lambda5007\) can be measured from photometry, finding that on average the EW derived from photometric excesses in filters is 30\% smaller than the direct spectroscopic measurement. We also discover that a stack of the line emitting galaxies shows a distinct morphology after subtracting imaging that contains only the continuum. This gives us a first view of the line or ionized gas emission from $z > 7$ galaxies, demonstrating that this material has a similar distribution, statistically, as the continuum. We also compare the derived SFRs and stellar masses for both parametric and non-parametric star formation histories, where we find that 35\% of our sample formed at least 30\% of their stellar mass in recent ($< 10$ Myr) starburst events.

\end{abstract}




\section{Introduction}

The high redshift universe is now being studied in depth by JWST as shown by a slew of papers on early galaxy discoveries in the past year  \citep{austin2023large, adams2023discovery, donnan2023evolution, finkelstein2023ceers, 2022ApJS..259...20H, 2022MNRAS.511.4464A, 2022ApJ...938L..15C, 2022NatAs...6..599D, trussler2023seeing, 2023MNRAS.523.1036B, 2023arXiv230414469M, 2023arXiv230800751F, 2023arXiv230810932C, 2022ApJ...940L..14N, curtis2022spectroscopic, 2023arXiv230602468H}. These studies have found that there are many more distant candidate galaxies at \(z > 7\) than inferred from before based on HST observations. However, uncovering their properties is really just in its infancy, and a major way to understand these systems is through spectroscopy. There are also many questions which we need to answer before we can reach the ultimate goal of using spectroscopy and imaging together to infer the physical properties of galaxies and therefore to determine galaxy evolution.  A major one is how well spectra and imaging agree in terms of deriving the physical properties of galaxies.

It is clear that spectroscopy with, in particular NIRSpec and also NIRCam/NIRISS in grism mode, are and will continue to be of major importance for the study of the first galaxies. At the same time, it will never be the case that we will obtain spectroscopy for all, or even a large fraction, of the most distant galaxies. The systems are too faint, and in many cases, too abundant to effectively obtain many spectra. Thus, we must resort to imaging, down to the completeness limit, to derive galaxy properties for understanding the galaxy population.  This is a well worn path and many papers have used imaging for the measurements of photometric redshifts, stellar masses, and derived star formation rates, amongst other properties \citep[e.g.,][]{adams2023discovery,austin2023large, fujimoto2023ceers, 2023arXiv230808540A}.

The purpose of this paper is therefore two-fold. We investigate how well we can derive properties of distant galaxies from their photometry by comparing the same properties as derived from spectroscopy. This includes a redshift comparison: $z_\text{Phot}$ vs. $z_\text{Spec}$, as well as measures of star formation rates and stellar masses.  For example, it might be the case that there is a systematic difference in the measurements of these quantities, such that the ones derived from photometry are for example lower than spectroscopy. If this is the case then we will need to account for this in future analyses.  We can also use spectroscopy and imaging together to derive unique properties of galaxies. An example of this is using the location of emission lines seen in spectroscopy which exist, and contribute flux, within various imaging filters. When this is well understood and well known (e.g., without uncertain redshifts) we can obtain an image of the line emission alone through subtracting filters that only contain continuum (no emission lines) from filters with flux arising from emission lines  \citep[][]{Hatch2013}.

This type of analysis has been carried out in other ways before, but never quite addressing the same questions we are here.  Previous similar work includes examining how well star formation and stellar masses can be measured based on comparisons with models and with different fitting codes and methods \citep[e.g.,][]{Mobasher2015, pacifici2023art}.  This is also the case for different photometric redshift codes \citep[][]{Dahlen2013}, where tests can be done to determine which methods and codes are the `best' for recovering correct photometric redshifts.  Recently this has been examined in terms of the stellar population properties of galaxies as derived through photometry, finding that stellar mass is consistent between different codes, although other properties derived from SED fitting can vary quite significantly \citep{pacifici2023art}. Here we examine similar questions, but we take a more detailed approach of comparing within the same code and same initial conditions how well the properties of galaxies can be derived based on photometry vs. spectroscopy. That is, we can determine the same features of galaxies using spectroscopic measurements, sometimes within the line emission detected, but otherwise fitting the spectrum. 

Thus, in this paper we investigate the spectroscopic properties of a sample of \(z > 7\) galaxies with reliable spectroscopic redshifts from NIRSpec on JWST within two different fields - CEERS \citep{finkelstein2023ceers} and JADES \citep{rieke2023jades, eisenstein2023overview}.  

The structure of this paper is outlined as follows. In Section~\ref{sec:data}, we detail the dataset sourced from the JADES and CEERS fields. Our main findings and analysis are presented in Section~\ref{sec:result}. A summary of our conclusions is provided in Section~\ref{sec:conclusions}. Throughout this work, we adhere to a standard cosmology with \(H_0=70\,\text{km}\,\text{s}^{-1}\,\text{Mpc}^{-1}\), \(\Omega_{\rm M}=0.3\) and \(\Omega_{\Lambda} = 0.7\) to facilitate comparison with other observational studies. All magnitudes reported are consistent with the AB magnitude system \citep{Oke1974,Oke1983}.

\section{Data and Reduction}
\label{sec:data}
The launch of the James Webb Space Telescope in December 2021 \citep{rigby2023science} provides an unprecedented opportunity to study the distant universe. Over the past year, several Cycle 1 observation programs have been conducted. In this paper, we analyze data from the JADES and CEERS programs, both in terms of imaging and spectroscopy.  Below we give some details of which data we use and how this data was reduced and processed.

\subsection{JADES NIRSpec Observations}
We use the first JADES released NIRSpec \citep{ferruit2022near} data (PI: Eisenstein, N. Lützgendorf, ID:1180, 1210), spanning the time-frame September 2022 to October 2022, with a focus on the publicly released data in GOODS-S field. The spectra are obtained through the application of both disperser/filter and PRISM/clear configurations. Specifically, the PRISM data covers 253 galaxies, and 198 of them have disperser/filter data. Four different disperser/filter combinations are used to acquire the spectroscopy: G140M/F070LP, G235M/F170LP, G395M/F290LP, and G395H/F290LP, with a wavelength coverage of $0.70 - 1.27 \mu$m, $1.66 - 3.07 \mu$m, $2.87 - 5.10 \mu$m, and $2.87 - 5.14 \mu$m, respectively. The three medium resolution filters have a nominal resolving power of R $\approx 1000$, while the high resolution data can reach R $\approx 2700$. In this paper, we primarily utilize the PRISM data, which covers a wavelength range of \(0.6 \, \mu\text{m}\) to \(5.3 \, \mu\text{m}\), and exhibits a spectral resolution of \( R \approx 30 - 330\) \citep{ji2022reconstructing}. 

Among the 253 observed galaxies, 13 are situated at \( z_{\text{spec}} > 7.0 \), with 11 of them having NIRCam observations. During these observations, three micro-shutters were activated for each target. An exposure protocol was implemented consisting of a three-point nodding sequence along the slit, with each nod lasting 8403 seconds, and the entire sequence repeated four times. This culminated in a total PRISM exposure time of up to 28 hours for some sources. The subsequent extraction of flux-calibrated spectra was carried out using specialized pipelines developed by both the ESA NIRSpec Science Operations Team and the NIRSpec GTO Team \citep{Bushouse2023}. A more detailed examination of the JADES/HST-DEEP spectra and the criteria used for sample selection is provided by \cite{eisenstein2023overview}.

\subsection{JADES NIRCam Observations}
The JADES NIRCam imaging observations \citep{rieke2023jades} cover both the GOODS-S and GOODS-N fields. In this paper, we focus on the GOODS-S field data (PI: Eisenstein, N. Lützgendorf, ID:1180, 1210). The observations utilise nine filter bands: F090W, F115W, F150W, F200W, F277W, F335M, F356W, F410M, and F444W, encompassing a spatial extent of 24.4 - 25.8 arcmin\(^2\). A minimum of six dither points was used for each observation, with exposure times spanning 14-60 ks. Correspondingly, the $5\sigma$ depths are within the range from 3.4 to 5.9 nJy, with flux aperture sizes varying between 1.26 and 1.52 arcsec. Across all filter bands, JADES ensures a high level of pixel diversity \citep{rieke2023jades}, thereby significantly reducing the impact of flat-field inaccuracies, cosmic ray interference, and other issues at the pixel level. In this paper, we utilize the publicly released JADES data and reductions.

\subsection{CEERS NIRSpec Observations}
The CEERS NIRSpec spectroscopic data \citep{fujimoto2023ceers, haro2023spectroscopic} were procured as part of the ERS program (PI: Steven L. Finkelstein, ID:1345). This dataset was designed to optimize the overlap with observations from both NIRCam and HST, using three medium resolution gratings \( R \approx 1000 \) and the PRISM \( R \approx 100 \). The PRISM data presented here are a reschedule of the original observations affected by an electrical short in CEERS Epoch 2 (December 2022). These rescheduled observations were executed in CEERS Epoch 3, February 2023. During this period, both NIRSpec pointings, namely NIRSpec11 and NIRSpec12, adhered to the standard CEERS MSA observational guidelines. Specifically, they encompassed three integrations with 14 groups in the NRSIRS2 readout mode per visit, leading to a total exposure time of 3107 s. Within these observations a trio of shutters was used to form slitlets, facilitating a three-point nodding sequence to enhance background subtraction. The PRISM disperser, ranging in wavelength from 0.6–5.3 $\mu$m, is characterized by its capacity to provide varied spectral details. In this paper, we use the NIRSpec data reduced by the Cosmic Dawn Center, which is published on the DAWN JWST Archive (DJA).\footnote{\href{https://dawn-cph.github.io/dja/blog/2023/07/18/nirspec-data-products/s p}{https://dawn-cph.github.io/dja/blog/2023/07/18/nirspec-data-products/}.} From this data set, there are 32 galaxies at $z_\text{spec} >7$, which we analyse in the following sections.

\subsection{CEERS NIRCam Imaging}
The CEERS (CEERS; ID=1345) NIRCam imaging \citep{bagley2023ceers} includes data across seven distinct filters: F115W, F150W, F200W, F277W, F356W, F410M, and F444W, with a $5\sigma$ depth of 28.6 AB magnitudes using 0.1 arcsec circular apertures.  The dataset encompasses observations collected during June 2022, accounting for 40\% of the total NIRCam area covered for CEERS in the latter half of the same year.

In this paper we utilise our own bespoke reduction of this data from the Cosmic Evolution Early Release Science Survey in the Extended Groth Strip field (EGS).  We have reduced this data independently ourselves using a custom set-up of the \textsc{JWST} pipeline version \textsc{1.6.2} using the in-flight calibration files available through the \textsc{CDRS 0942}.  We provide an extensive description of this process and the resulting data quality in \cite{2022ApJ...938L...2F, adams2023discovery}.

In parallel, v1.9 EGS mosaics HST data from the CEERS team are used. These are processed following the methodologies outlined in \cite{koekemoer2011candels}, which notably include enhancements in calibration and astrometric accuracy beyond what is available from the default HST archival pipeline, with a pixel scale of 0.03". For the HST data, two filters, namely F606W and F814W, are employed in our analyses due to their superior spatial resolution and depth when compared to HST/WFC3 images, and the fact that they are bluer than the JWST data.  We find that using these two HST filters within CEERS is critical for measuring accurate redshifts and other physical properties as this JWST dataset is missing the crucially important F090W band.


\subsection{Photometric Redshifts}
\label{photometric redshift method}
Analysing the quality and robustness of photometric redshift estimates is a key aspect of this paper, and thus we go into some detail in describing how they are measured here.  We use two different photometric redshift codes throughout this paper - \texttt{EAZY-PY} (hereafter \texttt{EAZY}) is our primary code, and then \texttt{LePhare} as a check on these values, both of which we describe below. Most of our results however are discussed mainly in terms of the \texttt{EAZY} code.

Our primary photometric redshifts arise from fitting our derived SEDs from the \texttt{EAZY} photometric redshift code \citep{brammer2008eazy}. This is the standard code used to measure photo-zs from the EPOCHS sample (\citealt{adams2023discovery}; Conselice et al. 2023, in prep). To carry out the photometric redshift analysis we use the BC03 template sets with a Chabrier initial mass function for our analyses, with details discussed in \cite{2003MNRAS.344.1000B} and \cite{2002ApJ...567..304C}, respectively. The templates we use include both exponential and constant star formation histories, whereby we use within these 10 characteristic timescales ranging from $0.01<\tau<13$~Gyr. In addition to this we use 57 different ages for the model galaxies spanning 0 to 13 Gyr. We include galaxies models which are at redshifts that range from $0<z<25$.  Dust is accounted for by using the prescription of \cite{calzetti2000dust}.  We allow for $E(B-V)$ values up to 3.5, to include any very dusty galaxies that may exist at these very high redshifts, and to determine the likely errors from low redshift contamination.   Our fitting of the photo-zs incorporates treatment for emission lines, and we apply the intergalactic medium (IGM) attenuation derived from \cite{1995ApJ...441...18M} when considering our fits.  The very blue templates we use are presented in \citet{larson2022spectral} as well as those which used by the JADES team \citep{2023arXiv230602468H}. These templates build upon the default template sets and incorporate galaxies that exhibit bluer colors and stronger emission lines, which are expected to be more appropriate for modelling the spectral energy distributions (SEDs) for those systems that are at $z > 7 $. 

In addition to \texttt{EAZY} we use photometric redshifts calculated with the \texttt{LePhare}  code.  The setup that we use is the same as we have used for the \texttt{EAZY} results described above. However, most of our results when using photometric redshifts arise from \texttt{EAZY}, and \texttt{LePhare} is used as a check on these. By utilizing multiple photometric redshift codes, we are able to cross-check the results for consistency and identify potential contaminants, thus ensuring the reliability of our final sample.

We do not use methods to fine-tune the zero points of the photometric bands, as the NIRCam modules consist of multiple individual chips (8 in the blue and 2 in the red), each with their own independent calibrations and photometric zero point offsets. Applying zero point modifications on a chip-by-chip basis, instead of on the final mosaic, would be necessary due to the small field of view covered by each chip, which results in a limited number of objects with spectroscopic redshifts within each chip, and leads to potential unnecessary biases determined by the positions of the galaxies in the NIRCam pointing. Doing this would also introduce potential biases towards systems with certain colors, which depend on the types of spectroscopically confirmed galaxies within each module. Discussions with members of the community have indicated that residual zero point errors were anticipated to be around 5 percent. Therefore, we have implemented a minimum 5 percent error on the measured photometry to account for potential zero point issues within the NIRCam reduction pipeline.

\section{Results}
\label{sec:result}

In this section we describe the basic results of our study by comparing photometric and spectroscopic data, and what can be learned by combining the two.  We include a comparison of the galaxy properties derived separately from the photometric and spectroscopic data, and how accurate we can derive properties from photometry by comparing with spectroscopy, assuming that the spectroscopic derivations are more accurate in some cases.   We later discuss the likelihood of this later case.

\subsection{Photometric vs. Spectroscopic Redshifts}

By far the most common way to estimate the distances of galaxies is through photometric redshifts. This is due to the fact that photometric redshifts can be measured when imaging is available for different galaxies in a variety of filters; this allows us to compare to templates of known redshifts and thus determine which is the best `fit'. In this section we carry out a comparison of how we measure the photometric redshifts for distant galaxies and how well these compare to the known high quality spectroscopic redshifts available from NIRSpec JWST data. 

There are however, two issues that we have to discuss concerning comparing the photometric and spectroscopic redshifts. The first is the selection of sources.  It is not enough to blindly measure photometric redshifts for everything that enters a catalogue, as the quality of those redshifts depends strongly on the quality of the data at all wavelengths, and how many filters a galaxy is detected within.

As described, the photometric redshift technique that we use to measure redshifts comes from \texttt{EAZY-PY} \citep{brammer2008eazy} and uses a variety of approaches discussed in Section 2.5.1. These methods and details of the photometric redshifts are further described in detail in \cite{2023MNRAS.518.4755A} and Conselice et al. (2023, in prep). For spectroscopic redshifts, we utilize data from the publicly available JADES catalog \citep{bunker2023jades}, as well as from the DAWN JWST Archive (DJA) for CEERS galaxies. We re-measure these spectroscopic redshifts ourselves using the [O~\textsc{iii}] \(\lambda5007\) line and find a good agreement with the published ones which we use throughout this paper.  For this initial comparison we just compare the photometric redshifts we obtain for all 43 galaxies in our sample (11 from JADES and 32 from CEERS), without consideration for whether these galaxies would be selected for observation based on other criteria, which we discuss in more detail below.

\begin{figure}
    \includegraphics[width=\columnwidth]{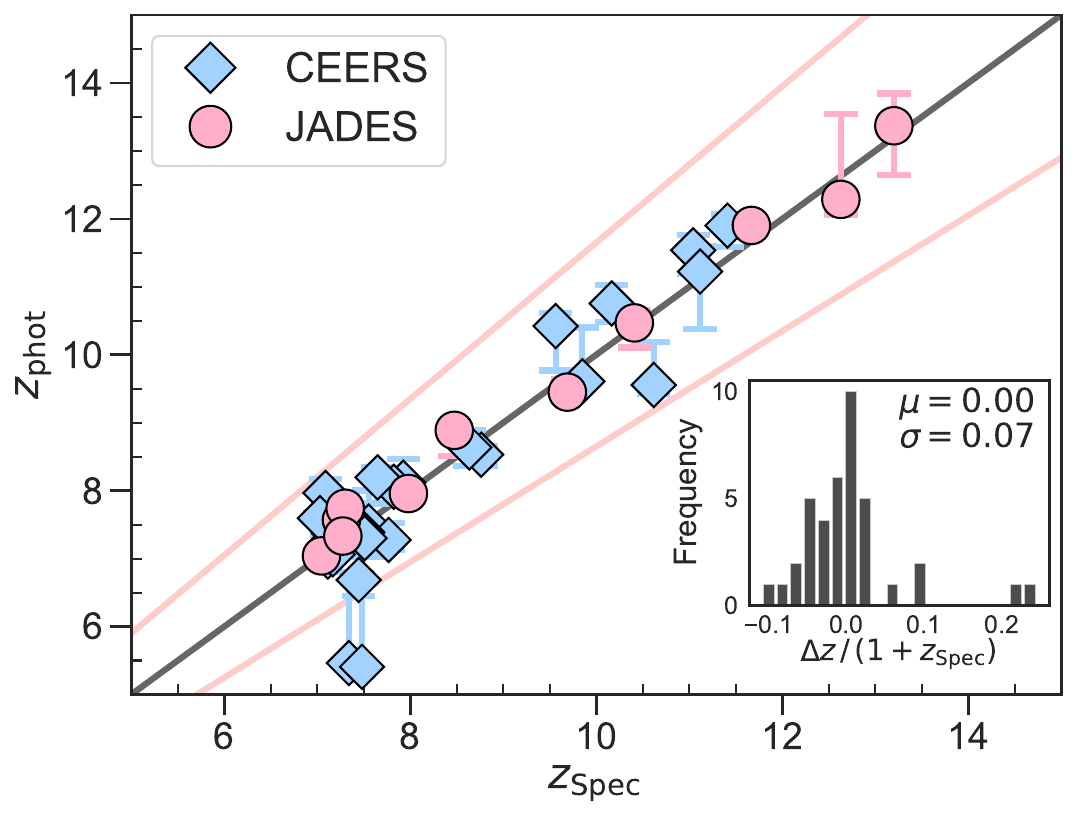}
    \caption{Comparison of spectroscopic redshifts with photometric redshifts from \texttt{EAZY}, for 11 JADES and 32 CEERS galaxies. A scatter plot between these values is presented and shown with an inset histogram at the bottom right corner, displaying the frequency distribution of the quantity \((z_{\text{spec}} - z_{\text{phot}})/(1 + z_{\text{spec}})\) of our galaxy samples.}
    \label{png: Combined_z_2}
\end{figure}

The outcomes of our redshift comparison are visually represented in \autoref{png: Combined_z_2}. We evaluate two statistical measures for all the galaxy samples: the outlier fraction \(\eta\) and the Normalised Median Absolute Deviation (NMAD). These two parameters are defined by the following expressions:
\begin{equation}
\eta = \frac{N_{115} + N_{85}}{N_{\text{total}}},
\label{eta}
\end{equation}
where \(N_{115}\) and \(N_{85}\) represent the counts of points lying above the line \(z_{\text{phot}} = 1.15 \times (z_{\text{spec}} + 1)\) and below the line \(z_{\text{phot}} = 0.85 \times (z_{\text{spec}} + 1)\), respectively. These counts indicate the presence of extreme outliers in the sample. The equation for calculating the NMAD is given by \citep[e.g.,][]{2019ApJ...876..110D}:
\begin{equation}
\text{NMAD} = 1.48 \times \text{med}\left|\frac{z_{\text{spec}} - z_{\text{phot}}}{1 + z_{\text{spec}}}\right|.
\label{NMAD}
\end{equation}

The values for these parameters, as applied to our data set, are detailed in \autoref{tab:eta_NMAD}. As is evident, our photometric redshift measurements show an exceptional concordance with the spectroscopically measured values. Notably, a mere 2.6\% of our samples qualify as extreme outliers in terms of their photometric redshifts. We find a very similar trend when using the \texttt{LePhare} photometric redshifts.  

We now would like to consider how the selection method we and others use in high redshifts papers would allow these galaxies to be correctly identified as high redshift (e.g., \citealt{adams2023discovery}; Conselice et al. 2023, in prep). The selection procedure in these papers, and others similar to them, uses more than just the best-fitting photo-z solution, including issues such as the limits on potential low-z solutions and the detection confidence of the photometry. In addition to having a high-z solution, these high-z papers often require that there be a low probability for the photometric redshift to be at lower-z. Another criteria for robust selection of high-redshift galaxies involves additional criteria, such as $> 3 \sigma$ detection in bands blueward of the Lyman break, a PDF integral of photometric redshifts between $\pm 0.1  z$ is greater than 60\% of the total, and $\chi^2$ values less than 6. These criteria are done to balance contamination with sample completeness. Thus we can test our methodology with this sample to see how many galaxies from this spectroscopic sample we would have included in our photometric samples in the EPOCHS papers.

In accordance with the selection criteria explained in our previous work (\citealt{2023MNRAS.518.4755A}; Conselice et al. 2023, in prep), 16 out of the 32 CEERS galaxies would be categorized as robust galaxies. The reasons that 16 galaxies would not have survived our selection are varied and depend on a few factors. Among the 16 galaxies that would make up this non-robust sample, 4 systems are excluded due to being near image edges or diffraction spikes. 1 galaxy is excluded for lacking observations in bands blueward of the Lyman break, and 11 are rejected owing to flux detections below \(5\sigma\) above the noise in the first, second, or both bands redward of the Lyman break. It is noteworthy that the CEERS team likely selected these 11 galaxies based on using smaller, $0.2 \, \text{arcsec}$ apertures for their photometry. Despite their faintness, our analysis still gets their redshifts correct. Thus, overall we only miss those galaxies which are too faint for reliable photometric redshifts or those that are in non-ideal regions of the images.

We generate both primary and secondary photometric redshift solutions for each galaxy in our study. The secondary redshift solutions are constrained to have a maximum allowable redshift of $z = 6$. In our robust galaxy samples, these secondary solutions typically exhibit an inferior fit quality compared to the primary solutions. This is substantiated by an average \(\Delta \chi^2\) value which is $\sim$ 35 higher than that of the primary solutions, for which the mean \(\chi^2\) is \(7.47\).

\begin{table}
\centering
\caption{Values of $\eta$ and NMAD for JADES, CEERS, and Joint Data. The outlier fraction, $\eta$, expressed as a percentage and defined by \autoref{eta}, measures the proportion of extreme outliers in the redshift comparison. The NMAD, calculated using equation \autoref{NMAD}, estimates the scatter in the redshift differences, adjusted for scale. The low values of both metrics attest to the accurate measurement of our photometric redshifts.}

\label{tab:eta_NMAD}
\begin{tabular}{cccc}
\hline
Parameters   & JADES  & CEERS   & Joint  \\
\hline
$\eta$  & 0.0 \%   & 3.6 \%     & 2.6 \%   \\
NMAD         & 0.027  & 0.036   & 0.035  \\
\hline
\end{tabular}
\end{table}

\subsection{Galaxy Intrinsic Properties}
\label{sec: Galaxy Intrinsic Properties}
We analyze the intrinsic properties of our galaxy samples, both spectroscopically and photometrically, using various methodologies. Specifically, we use \texttt{Bagpipes} \citep{carnall2018inferring} to fit the photometric and spectroscopic galaxy data separately with six parametric SFH models—log-normal, delayed, constant, exponential, double delayed, and delayed burst—along with a non-parametric Continuity model \citep{leja2019measure}, while fixing the redshift to the spectroscopic redshift in both cases. We use Log10 priors for dust, metallicity, and age. The reason for selecting Log10 priors is because we expect high redshifts galaxies to be young, with lower metallicity and for being less dusty. We set prior limits for metallicity in the range of $[1\text{e-06}, 10.0] \, \text{Z}_{\odot}$, dust prior in the range of $[0.0001, 10.0]$ in $A_\text{V}$, the time assumed for star formation to start at 0.001 Gyr, the time assumed for star formation to stop at $t_\text{U}$, with $t_\text{U}$ denoting the age of the Universe. In addition, \cite{kroupa2001MNRAS.322..231K} IMF, \cite{2003MNRAS.344.1000B} SPS model, and the  \cite{calzetti2000dust} dust attenuation model is implemented.  For each model, we examine no effects other than using different SFR timescales — 5 Myr, 10 Myr, and 100 Myr — on the derived properties. These timescales only impact the measured SFR.

Since there are no significant differences in galaxy parameters derived from various models, we have chosen to focus our analysis on the results obtained using the log-normal SFH model. For each property computed from \texttt{Bagpipes}, the derived values are represented by the median of their respective PDF. The lower and upper uncertainties are determined as the differences between the 50th percentile and the 16th, and between the 84th and the 50th percentiles, respectively.


In our spectroscopic fitting, we incorporate three additional considerations \citep{carnall2019vandels}: velocity dispersion, flux calibrations, and noise. The velocity dispersion is modelled by setting the width of the Gaussian kernel in velocity space to be convolved with the spectroscopic output, within a range of $[1, 1000]$ km/s. For flux calibrations, we address potential discrepancies between photometric and spectroscopic measurements by fitting a Chebyshev polynomial perturbation to the spectroscopic data \citep{carnall2019vandels}. This method assists in correcting calibration issues and aligning the models. To account for noise, we introduce a factor that applies a multiplicative adjustment to all spectroscopic uncertainties. Moreover, to evaluate potential slit losses, we simulate photometric flux using the observed spectral data. Our analysis reveals a maximum discrepancy of $\sim$20\% between the observed photometric flux points and the simulated data, predominantly in the NIRCam filter F090W. This discrepancy is likely attributed to the fact that this band is blueward of the Lyman break for our sample galaxies at redshifts \(z > 7\), resulting in a significant drop in flux. Consequently, the noise dominates in this band. For other filter bands, no discernible differences are observed.


We produce a scatter plot with photometrically-derived values on the y-axis and spectroscopically-derived values on the x-axis, for \texttt{Bagpipes} derived stellar masses, formed masses, SFRs, and dust extinction values (A$_\text{V}$). Using the Bayesian Markov Chain Monte Carlo (MCMC) method, we compute the line of best fit for each plot via the \texttt{emcee} package \citep{foreman2013emcee}. Specifically, we employ 100,000 steps and 50 walkers to generate candidate gradients and y-intercept values. For both sets of values, we adopt the mean as the representative value and use the 1$\sigma$ deviation as the associated uncertainty, as the distributions follow a perfect Gaussian. In addition, the Pearson correlation coefficient between the spectroscopic and photometrically derived values is determined, and its uncertainty is calculated using the Fisher transformation. Specifically, the Pearson correlation coefficient \( r \) is transformed into a \( z \)-score using the Fisher transformation, which is given by 
\( z = \frac{1}{2} \ln \left( \frac{1 + r}{1 - r} \right) \).
This transformation ensures that the distribution of \( z \) is approximately normal. Once \( z \) is obtained, the 95\% confidence interval for it is calculated. Subsequently, this confidence interval is transformed back to the correlation coefficient scale using the inverse Fisher transformation, represented by
\( r = \frac{e^{2z} - 1}{e^{2z} + 1} \). 
Thus, providing the 95\% confidence interval for the original correlation coefficient \( r \). The results of gradients, intercepts, and correlation coefficients using 100 Myr SFR timescale are presented in \autoref{tab:bagpipes_regression_correlation}.

\begin{table*}
    \centering
    \caption{Linear regression and Pearson correlation analysis between spectroscopic and photometric results for different galaxy properties derived from \texttt{Bagpipes} using 100 Myr SFR timescale. The gradient and y-intercept of the regression model are computed, and the uncertainty in the correlation coefficient is calculated using the Fisher transformation. The 1$\sigma$ values (or scatter) for residuals between the best fit line and scatter points are shown.}

    \label{tab:bagpipes_regression_correlation}
    \renewcommand{\arraystretch}{1.4}
    \begin{tabular}{lccccl}
        \hline
        Property & Correlation & Gradient & Intercept & Residual $1\sigma$\\
        \hline
        Stellar Mass [log$_{10}(\text{M}_{\odot})$] & $0.62^{+0.39}_{-0.28}$ &$0.55 \pm 0.11 $& $3.49 \pm 0.90$ & 0.37 [log$_{10}(\text{M}_{\odot})$]\\
        Mass Formed [log$_{10}(\text{M}_{\odot})$] & $0.58^{+0.34}_{-0.35}$ & $0.53 \pm 0.12$ & $3.72 \pm 0.95$ & 0.41 [log$_{10}(\text{M}_{\odot})$]\\
        Star Formation Rate [$\text{M}_{\odot} \, \mathrm{yr}^{-1}$] & $0.64^{+0.15}_{-0.22}$ & $0.67 \pm 0.13$ & $0.42 \pm 0.30$&1.37 [$\text{M}_{\odot} \, \mathrm{yr}^{-1}$]\\
        Dust Extinction (Av) [AB mag] & $0.61^{+0.16}_{-0.24}$ & $0.49 \pm 0.1$& $0.15 \pm  0.04$ & 0.20 [AB mags]\\
        \hline
    \end{tabular}
\end{table*}
\subsubsection{Quality of the \texttt{Bagpipes} fits}

In this short section we discuss how well we can fit the SEDs of our galaxies with the \texttt{Bagpipes} fits and the underlying models which we use. These are standard models which have been used throughout the literature for years, but it might be the case that at these higher redshifts galaxy SEDs might be better fit by, for example, models in which the IMF differs from the assumption (perhaps top-heavy) or by models which incorporate binary stars (e.g., BPASS) \citep[e.g.,][]{Eldridge2009}.  One way to determine this is through examining how well our SEDs are fit by these models as determined through the $\chi^2_\text{reduced}$ values of these fits.

We evaluate the goodness of fit for our models by calculating the \(\chi^2_\text{reduced}\) for both photometric and spectroscopic fitting. Both our JADES and CEERS samples exhibit comparable photometric \(\chi^2_\text{reduced}\) values, meaning that there is not one particular sample which is better fit by our methods than the other.   Precisely, the mean photometric \(\chi^2_\text{reduced}\) for the JADES samples is \(1.5 \pm 0.6\), whereas for CEERS samples, it stands at \(2.0 \pm 1.1\). This indicates a similar and good level of photometric fitting quality for these two sets of galaxy samples.

However, the spectroscopic fitting quality for CEERS samples appears to be slightly inferior based on this statistic. The mean $\chi^2_\text{reduced}$ for JADES is $1.54 \pm 0.65$. In contrast, the corresponding value for the CEERS sample rises to $3.19 \pm 1.34$, nearly double that of JADES. We speculate that the worse fitting quality for CEERS is primarily attributed to its shorter exposure time. Some JADES galaxies have exposure times extending up to 28 hours, whereas CEERS employs an exposure time of less than an hour.  Whilst the larger errors on the fainter observations should account for this, it is possible that these are being underestimated in our fits, and therefore resulting in higher $\chi^2_\text{reduced}$ values. In any case, we do not observe large $\chi^2_\text{reduced}$ values that would suggest the models we fit are inherently flawed. However, a more detailed analysis is warranted and necessary, but this is beyond the scope of this paper.


\subsubsection{Measuring Galaxy Stellar masses}
In this section, we examine the various different ways in which stellar mass and formed mass are derived from \texttt{Bagpipes} using the spectroscopic and photometric data. Stellar mass represents the present-day mass of the galaxy, while the formed mass incorporates the observed mass plus the return mass, accounting for the mass from exploded stars that contribute to the formation of new stars. Consequently, the formed mass is always greater than the stellar mass.  Also, the stellar mass is the only quantity we can directly compare with given that this is what we are observing. In addition, different SFR timescales dictate the duration over which the star formation rate is averaged, and these do not influence the derived galaxy masses. Thus, we present only the 100 Myr averaged SFR results here. In \autoref{tab:bagpipes_regression_correlation}, we show the correlation coefficient and the parameters of the best-fit line for the spectroscopically and photometrically derived values of these two quantities. Generally speaking, these two masses derived from both methods are in moderate agreement, with high scatter. 

\begin{figure}
	\includegraphics[width=\columnwidth]{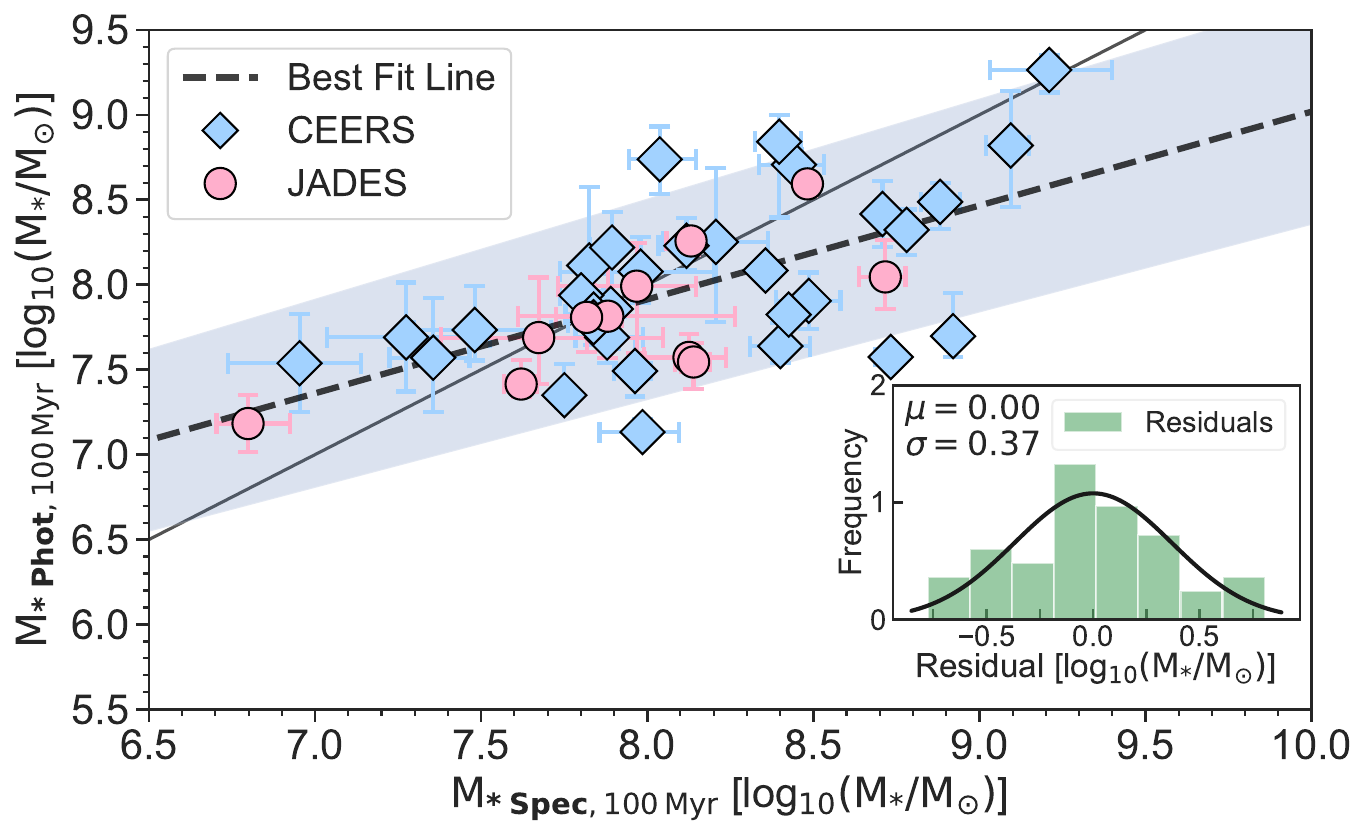}
    \caption{The comparison of galaxy stellar masses derived from spectroscopic and photometric data using \texttt{Bagpipes} fitting, based on a log-normal star formation history with a 100 Myr SFR timescale. The best fit line for all data points has a gradient of $0.55 \pm 0.11$, and an interception of $3.49 \pm 0.90$ \(\log_{10}(\text{M}_* / \text{M}_{\odot})\), as shown by the dashed line.  The solid line shows the 1:1 relation between the two masses.  The correlation coefficient between the spectroscopic and photometric measurements is $0.62^{+0.39}_{-0.28}$.  In general, we find a better agreement between these methods of measuring stellar masses at intermediate masses. At the lowest masses the photometric method gives larger masses, whereas at the higher masses the spectroscopic measurement of stellar mass is larger. }
    \label{png:Mass_all}
\end{figure}

We present a graphical comparison for stellar masses in \autoref{png:Mass_all}. For our galaxy samples, the stellar masses for both CEERS and JADES range from \(\log_{10}(\text{M}_* / \text{M}_{\odot}) = 6.8\) to 9.3, with individual means of 8.0 for both fields, consistent with the findings of \cite{fujimoto2023ceers}. The correlation coefficient for spectroscopically and photometrically derived stellar masses is \(0.62^{+0.39}_{-0.28}\), which indicates moderate agreement between these two methods. However, the \(1\sigma\) residual of 0.37 \(\log_{10}(\text{M}_* / \text{M}_{\odot})\) from the best fit line suggests high scatter in the data. We hypothesize that this scatter arises from some photometric bands being affected by strong emission lines of H\(\beta\) and [O~\textsc{iii}], thereby reducing the accuracy of the stellar masses. Further investigation reveals that galaxies with this pronounced scatter generally exhibit high star formation rates. Although there isn't a universally strong agreement across all mass ranges, a notably better alignment is observed within the mass range \(\log_{10}(\text{M}_* / \text{M}_{\odot}) = [7.6, 8.2]\).

\subsubsection{Measuring Star Formation Rates}
\label{sec:SFR}
In this section, we employ three methods to measure the SFR: which we name as: \texttt{Bagpipes}, UV luminosity, and H$\beta$ line luminosity.
For the \texttt{Bagpipes} method, we not only analyze the correlation in SFR derived both photometrically and spectroscopically, but also study the variations in the derived SFR values when employing different timescales: 100 Myr, 10 Myr, and 5 Myr. We use this to investigate the star formation history of our sample and to determine when the stellar masses of these galaxies formed. We then compare these SFR measurements with those from direct line and UV measures. The specific parameters for the \texttt{Bagpipes} fitting are detailed in Section \ref{sec: Galaxy Intrinsic Properties}.

Beyond the insights provided by the \texttt{Bagpipes} method, we further measure the SFR directly using H$\beta$ line luminosity from spectrum, and UV luminosity derived from the photometry. Each technique, as elaborated in this section, calculates the SFR over distinct timescales. For instance, the Hydrogen $\beta$ method predominantly captures recent SFRs—about 10 Myr prior to observations. In contrast, the UV luminosity method gauges the SFR over a longer window, specifically the $\sim 100$ Myr preceding the observations \citep{kennicutt2012star}.

For SFRs measured from the H$\beta$ line, we employ the calibration proposed by \cite{kennicutt2012star}. The approach harnesses synthetic stellar populations and SEDs to calibrate various SFR tracers, relying on a standard IMF for enhanced results over previous calibrations. Typically, the H$\alpha$ luminosity is used for SFR calculations due to its direct relationship with recent star formation, and this relationship is expressed as:
\begin{equation}
\log\dot{M_{*}} (\text{M}_{\odot} \text{yr}^{-1}) = \log L_{\text{H}\alpha} - \log C_{\text{H}\alpha},
\label{eq:SFR UV}
\end{equation}
where \( L_{\text{H}\alpha} \) is the H$\alpha$ luminosity and \( C_{\text{H}\alpha} \) is the calibration constant with \( \log C_{\text{H}\alpha} = 41.27 \). However, in our high-redshift galaxy samples, the H$\alpha$ line is redshifted beyond the NIRSpec wavelength range. We therefore measure the H$\beta$ line luminosity, from which we derive the H$\alpha$ luminosity using the ratio $L_{\text{H}\alpha} / L_{\text{H}\beta} = 2.86$, applicable in dust-free star-forming regions \citep{kennicutt2012star}.  We later discuss how viable this assumption is and how it might influence our measurements. 

To estimate SFRs directly from the photometry, we employ the conversion from the UV luminosity directly measured \( L_{\text{UV}} \) to SFR as presented in Equation \ref{eq:SFR_UV}. In this case we do correct for dust obscuration by measuring the rest-frame UV using a technique that involves utilising the UV $\beta$ slope.  We fit a power law to the rest-frame UV photometry of the galaxy to determine the proportionality constant, $\beta$. The dust corrected SFR in solar mass per year is then computed using the  equation from \cite{madau2014cosmic}:
\begin{equation}
\mathrm{SFR}_{\mathrm{UV}} = \kappa \cdot L_{\mathrm{UV}} \cdot 10^{0.4 (4.43 + 1.99 \beta)},
\label{eq:SFR_UV}
\end{equation}
where \(\kappa = 1.15 \times 10^{-28}\, \mathrm{M}_{\odot}\, \mathrm{yr}^{-1}\, \mathrm{erg}^{-1}\, \mathrm{s}\, \mathrm{Hz}\), is the proportionality constant that accounts for the efficiency of star formation and the IMF \citep{salpeter1955luminosity}, $4.43 + 1.99 \beta$ is the dust correction factor \(A_\text{UV}\) \citep{1999ApJ...521...64M}, and \(L_{\mathrm{UV}}\) is the UV luminosity of the galaxy.   We use these star formation calibrations and measurements in the following subsections.

\subsubsection{Photometry vs. Spectroscopy SFRs}

\begin{figure}	\includegraphics[width=\columnwidth]{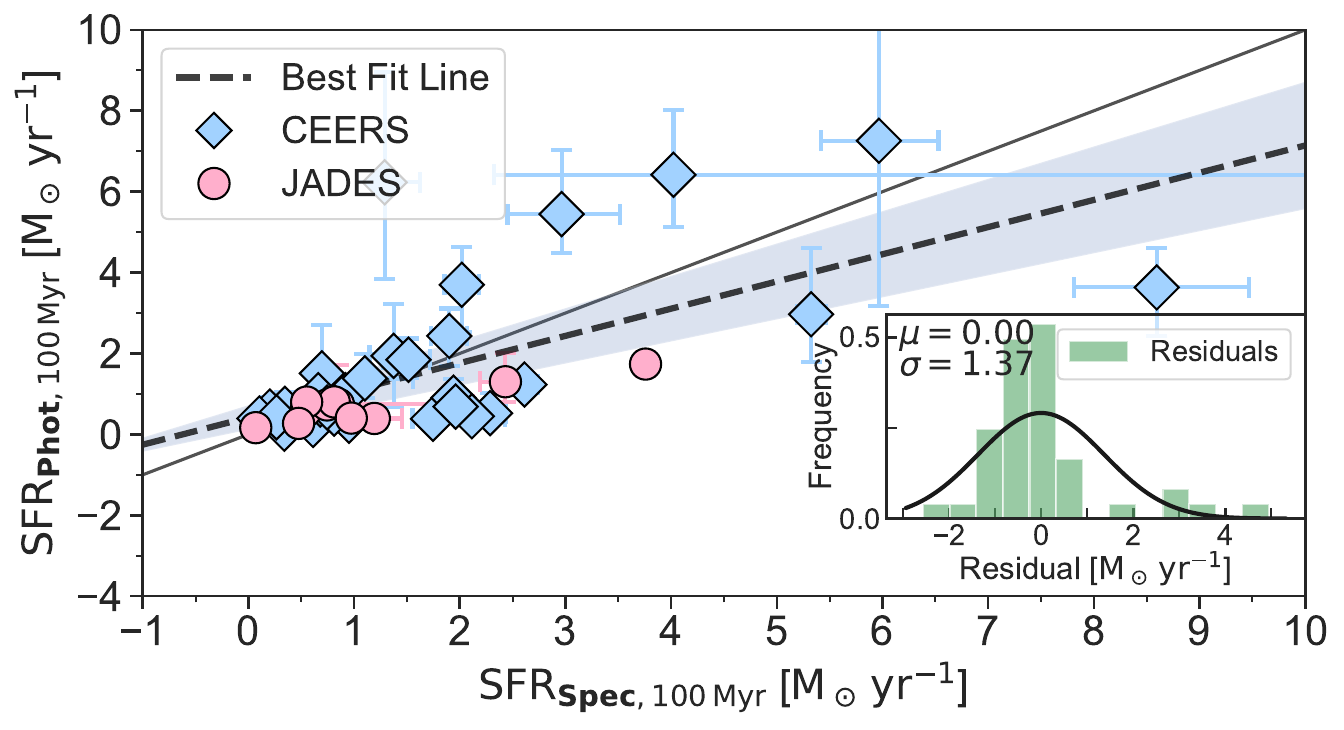}
    \caption{Comparison of SFR derived from spectroscopic and photometric \texttt{Bagpipes} fitting for 43 galaxies using a 100 Myr SFR timescale.  That is the measured SFR using the same way with the same code, but one axis shows the photometric values while the other the spectroscopic results. We find a correlation coefficient of $0.64^{+0.15}_{-0.22}$ and that 80\% of the galaxies have SFRs in the range $[0.3, 3] \, \mathrm{M}_{\odot} \, \mathrm{yr}^{-1}$.}
    \label{Bagpipes SFR}
\end{figure}

In this subsection we investigate how well fits to spectroscopy compare with fits to the photometry for measuring star formation rates within our sample of galaxies. The reason for doing this is to determine how well we can measure the SFR in terms of internal consistency, but also if we assume that the star formation rate measured from spectroscopy is somehow more 'correct' than with photometry, how different these two measures would be. In \autoref{Bagpipes SFR} we show a comparison of \texttt{Bagpipes} derived spectroscopic and photometric SFR using a 100 Myr timescale. In our sample, the majority of galaxies exhibit an SFR ranging from $\sim 0.3$ to $\sim 3 \, \text{M}_\odot \text{yr}^{-1}$, with a number of systems having higher SFRs, reaching up to $\sim 9 \, \text{M}_\odot \text{yr}^{-1}$. We find that the JADES sources with NIRSpec data typically exhibit a lower mean SFR of \(1.6 \, \text{M}_\odot \text{yr}^{-1}\), compared to those from the CEERS field which have a mean SFR of \(5.6 \, \text{M}_\odot \text{yr}^{-1}\).  However, it is important to note that this is within the errors of these measurements. These differences underline the significance of selection biases in studying diverse high-redshift galaxies, emphasizing the need for a more comprehensive spectroscopic approach in future endeavors. 

Furthermore, the correlation coefficient between these star formation measurements is $0.64^{+0.15}_{-0.22}$, signifying a good agreement between the two methods. Notably, there is an especially strong concordance between photometrically and spectroscopically derived SFRs, for SFR values up to $2$ $\text{M}_\odot \text{yr}^{-1}$. It is only at the higher end of the star formation where we find that the photometry is higher. However, it is important to keep in mind that these differences are at about the level of the uncertainty in these values.

\subsubsection{Bursty Star Formation Events}
In this section, we present three ways in which the bursty SFH nature for our galaxy sample is identified and verified within these high redshift galaxies. We are able to do this as we have the ability to determine the SFR accurately knowing the correct redshift of our systems.

\begin{figure}
	\includegraphics[width=\columnwidth]{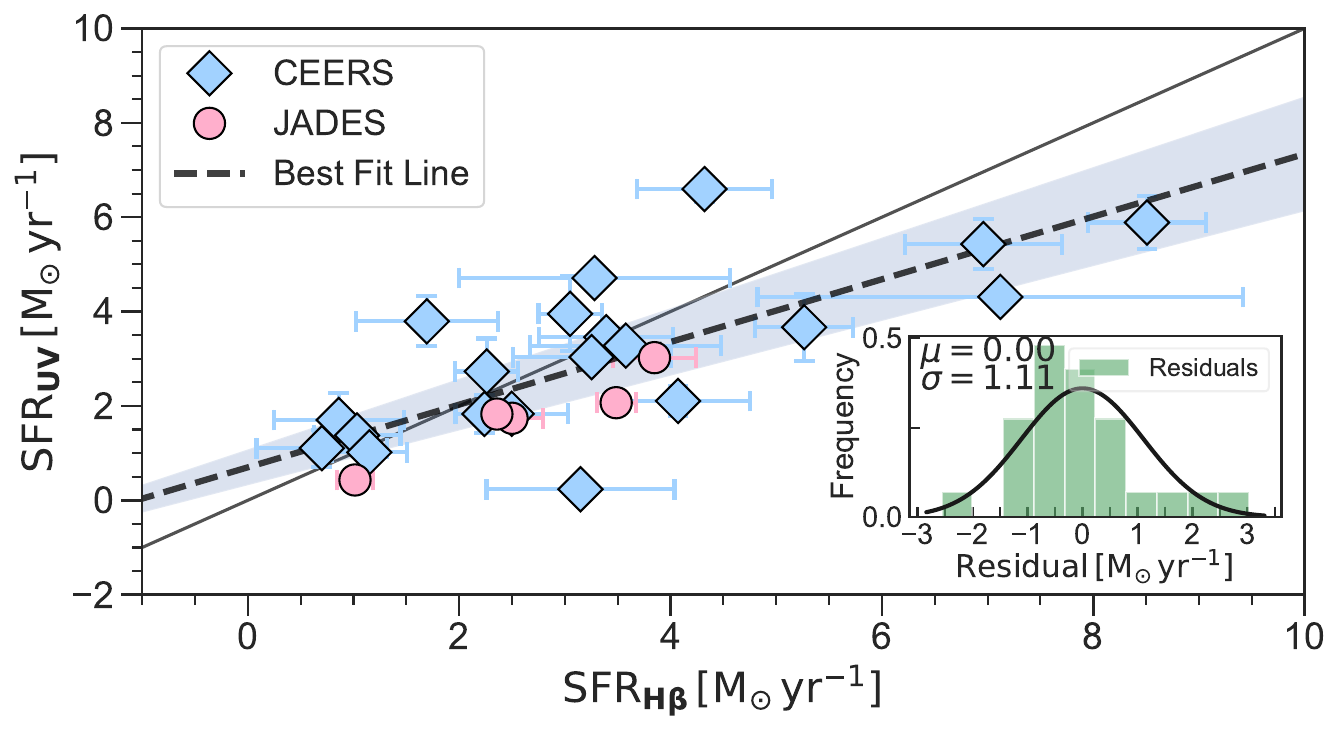}
    \caption{Comparison of SFR derived from H$\beta$ lines and UV luminosity for 25 galaxies exhibiting H$\beta$ lines under dust free assumption. 68 \% galaxies have higher H$\beta$ derived SFR values than that obtained using the UV luminosity method, with a factor up to $2.4 - 13.5$.}
    \label{fig:SFR UV Halpha}
\end{figure}

Firstly, we turn our attention to the comparison between SFR derived from H$\beta$ line emission and the UV luminosity. Of our samples, 5 out of 11 JADES galaxies and 20 out of 32 CEERS galaxies exhibit an H$\beta$ line which we can measure. The comparison for these galaxies is illustrated in \autoref{fig:SFR UV Halpha}. Given that the H$\beta$ results have not been corrected for dust, we opted for a consistent comparison by assuming a dust-free condition for the UV-derived SFR as well.  As these are low mass high redshift galaxies, they are unlikely to be very dusty in any case.  Consequently, the term $10^{0.4 (4.43 + 1.99 \beta)}$ as outlined in \autoref{eq:SFR UV} is omitted from these comparisons. 

It is worth noting that if dust correction is taken into account, then the effect is stronger in the rest-frame UV than in the rest-frame optical where H$\beta$ is located. Upon analysis, 60\% of the CEERS samples show a higher SFR from the H$\beta$ line luminosity measurement compared to that from the UV luminosity, while this observation is true for all the JADES samples. The SFR derived from H$\beta$ line luminosity can be as much as 2.4 times higher for JADES samples and 13.5 times for CEERS samples, which may well be due to photometric selection biases in the way these galaxies are selected.  

The higher SFR from the H$\beta$ line method most likely arises from the differing timescales each method probes. The UV luminosity reflects the SFR over the previous 100 Myr, while H$\beta$ traces the SFR over much shorter timescales of $\sim$10 Myr. Such findings suggest a bursty phase of star formation in these galaxies over the recent few million years (see below for further proof of this). One factor that may bias the sample towards higher SFR during the past 10 Myr is that we are only showing the H$\beta$ SFRs for galaxies with an identifiable H$\beta$ detection. Another issue which we have ignored in this calculation is the dust content. It might be the case that the dust extinction is high enough to attenuate the UV light more than the H$\beta$ line flux such that it only appears to be lower.  We investigate the dust in more detail in Section \ref{sec: dust attenuation}, however we give some indication for its impact here.   Using the \cite{calzetti2000dust} dust law we find an attenuation of A$_\mathrm{UV} = 0.25$, A$_{\mathrm{H}\beta} = 0.13$ for our galaxies.  This leads to a relative increase in UV SFR over H$\beta$ by about 10\% (25\% increase in UV vs. a 12\% increase in H$\beta$ flux) which is not nearly enough to create UV star formation rates that match the observed H$\beta$. Thus, we can conclude that there is an intrinsic difference in what these two star formation rates are measuring.

To investigate the bursty nature of the SFH of these galaxies more thoroughly, we utilize the non-parametric 'Continuity' model presented by \cite{leja2019measure}. Our analyses yield consistent findings: galaxies with higher H$\beta$-derived SFR do indeed exhibit a notable burst in their SFH when interpreted through the Continuity model. Specifically, for a majority of these cases, the timing of these star formation bursts is identified to occur within a timeframe spanning $0.3$ to $0.7$ Gyr.

Another aspect that indicates a bursty SFH is from the specific star formation rate, defined as 
\begin{equation}
    \text{sSFR} = \frac{\text{M}_\text{formed}(<t) / t}{\text{M}_*}, 
\end{equation}
\noindent where \( \text{M}_\text{formed} \) represents the mass formed within the past \( t \) years, and \( \text{M}_* \) is the observed stellar mass of the galaxy. If a galaxy formed all its mass within the past \( t \) years, then \( \text{M}_\text{formed} = \text{M}_* \), neglecting any stellar mass loss through stellar evolution processes, resulting in a maximum sSFR of \( \text{sSFR} = (1 / t) \).  

\begin{figure}

	\includegraphics[width=\columnwidth]{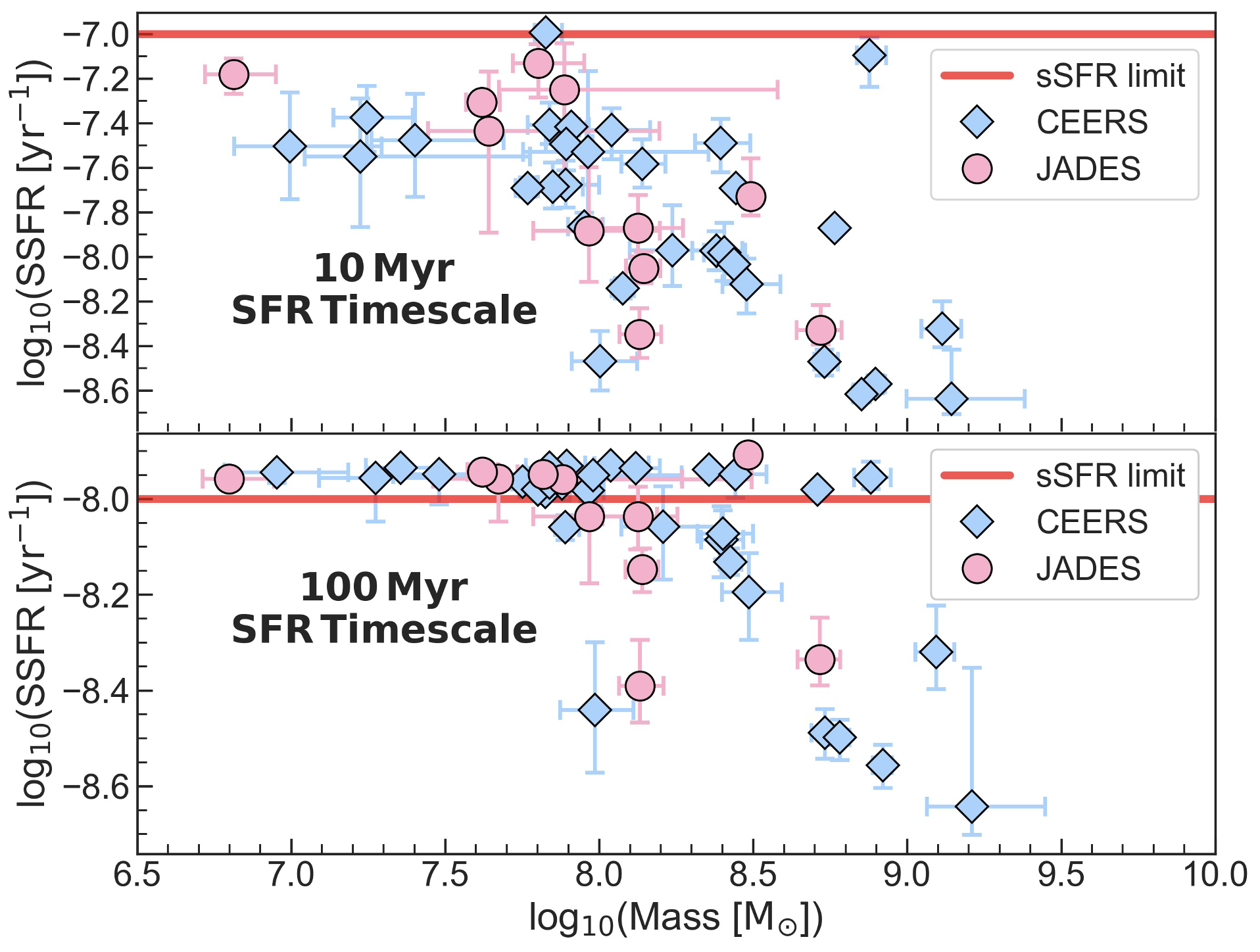}
    \caption{Comparison of $\log_{10}(\text{sSFR})$ derived from \texttt{Bagpipes} spectroscopic fitting for two distinct SFR timescales: 10 Myr (top) and 100 Myr (bottom). In the 100 Myr SFR timescale graph, 61\% of CEERS galaxies and 73\% of JADES galaxies reach the $\log_{10}(\text{sSFR}) = -8$ limit, suggesting these galaxies formed their entire stellar mass within the past 100 Myr. Conversely, in the 10 Myr graph, only 2 galaxies from both CEERS and JADES achieve $\log_{10}(\text{sSFR}) = -7$ limit. This evidence advocates for the appropriateness of a 10 Myr timescale over a 100 Myr timescale, especially in light of bursty star formation patterns observed in recent million years.}
    \label{fig: ssfr}
\end{figure}

We utilise \texttt{Bagpipes} to derive the sSFR both spectroscopically and photometrically. The majority of our sample display higher values of photometrically derived \( \log_{10}(\text{sSFR}) \) compared to the spectroscopically derived values, with the most significant discrepancy being 11\% observed in both the CEERS and JADES samples. Furthermore, in \autoref{fig: ssfr}, we show \( \log_{10}(\text{sSFR}) \) for our samples derived from \texttt{Bagpipes} spectroscopic fitting under 10 Myr and 100 Myr SFR timescale.  Most galaxies attain the maximum \( \log_{10}(\text{sSFR}) \) value of \( \log_{10}(1/t) = -8 \) using a 100 Myr SFR timescale.  This implies that most galaxies are consistent with forming most of their stars within the past 100 Myr.  Additionally, we also find $45\% \pm 20\%$ of JADES galaxies and $34\% \pm 11\%$ of CEERS galaxies formed 30\% of their total mass within the past 10 Myr. In addition, from this 10 Myr timescale model, two CEERS samples achieve a \( \log_{10}(\text{sSFR}) \) value of \( -7 \), suggesting they formed their entire stellar mass within this period, while two JADES galaxies reach \( -7.2 \), indicating approximately 60\% of their stellar mass was formed during the past 10 Myr, both signifying periods of intense star formation. These observations underscore the bursty nature of star formation in the last few million years for these galaxies. A comparative analysis using a 5 Myr SFR timescale does not produce results significantly different in sSFR from those obtained with a 10 Myr SFR timescale, indicating a relatively stable star formation rate across these two timescales.

\subsubsection{Dust attenuation and Star Formation Rate}
\label{sec: dust attenuation}

It is crucial to emphasize the dust correction factor used in our SFR measurements across different methodologies, as well as when we do and do not use it. As mentioned, we employ three methods to derive SFR measures: this includes the measurements from the \texttt{Bagpipes} code, SFR derived from UV luminosity with dust corrections, and SFR calculated from the H$\beta$ line luminosity. Dust attenuation effects are only considered for the SFR derived using the first two methods, assuming a \citep{calzetti2000dust} dust attenuation model. 

Specifically, \texttt{Bagpipes} fitting applies this dust law to derive \(A_{\text{V}}\), representing the attenuation in the V-band. In contrast, when calculating SFRs via the UV luminosity method, we utilize the same dust model but determine \(A_{\text{UV}}\) using the formula by \cite{1999ApJ...521...64M}: \(A_{\text{UV}} = 4.43 + 1.99\beta\).   Essentially this formula allows us to determine the dust extinction in the UV by measuring the UV slope $\beta$, which we do using methods outlined in \cite{austin2023large}.  It is worth noting that the dust law from \cite{1999ApJ...521...64M} is primarily tailored for $z\sim4$ galaxies and thus may not be directly applicable for our sample at $z>7$. We chose to use it in the absence of a currently widely accepted dust attenuation law for high-redshift galaxies. A comprehensive discussion regarding this choice can be found in Austin et al. (in prep).  One way that we can see this problem is that some of the values for the $A_{\text{UV}}$ are actually negative using this method, which is meaningless in this context. 

Considering the potential unsuitability of the \cite{1999ApJ...521...64M} dust attenuation relation for our high-redshift samples, it is essential to gauge its influence. We address this by comparing the dust-corrected SFR values derived from UV luminosity with those derived from \texttt{Bagpipes}. As the UV luminosity method computes the SFR over an approximate 100 Myr timescale, assuming a lognormal SFH, we adhere to the same parameters in our \texttt{Bagpipes} fits. In addition, while both methodologies employ the \cite{calzetti2000dust} dust model, their applications differ. The UV luminosity method calculates the attenuation \(A_{\text{UV}}\) in the UV band, whereas \texttt{Bagpipes} determines the attenuation \(A_{\text{V}}\) in the V band. To ensure a consistent comparison we convert the UV luminosity's dust correction factor from \(A_{\text{UV}}\) to \(A_{\text{V}}\), accounting for the discrepancies in dust attenuation between the two methods. This conversion leverages the relationship \(A_{\text{UV}} / A_{\text{V}} = S\), with \(\log_{10}{S} = 0.40\) \citep{2020ARA&A..58..529S}. 

\begin{figure}
	\includegraphics[width=\columnwidth]{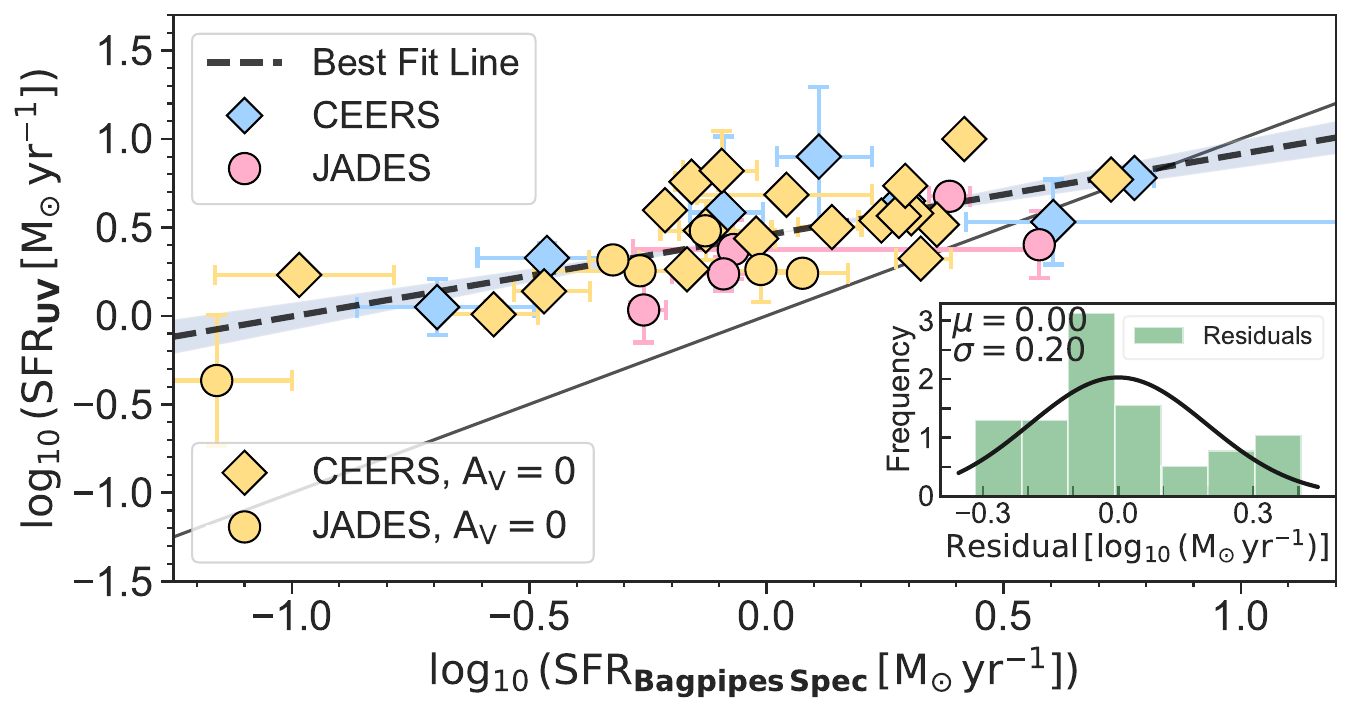}
     \caption{Comparison of the SFR determined from UV luminosity to those derived via \texttt{Bagpipes} spectroscopic fitting, using log-normal SFH and a 100 Myr SFR timescale. We convert the UV luminosity dust attenuation from \(A_{\text{UV}}\) to \(A_{\text{V}}\) to ensure consistency in the dust attenuation factor with \texttt{Bagpipes}. Yellow points represent galaxies with negative \(A_{\text{UV}}\), which is physically meaningless and are thus set to $=0$ (dust-free). Overall, the correlation coefficient is $\sim 0.7$, although an ideal correlation would yield a value of 1. This discrepancy stems from the erroneous $A_\text{UV}$ value we calculated, using the scaling relation from \protect\cite{1999ApJ...521...64M}, which is only applicable at lower redshifts ($z \sim 4$).}

    \label{fig:SFR UV Bagpipes}
\end{figure}

After these SFR values measured using both methods are aligned in terms of the same dust correction factor ($A_{\text{V}}$), we can assess the potential discrepancies between the two. The comparison of SFR derived from these two methods is shown in \autoref{fig:SFR UV Bagpipes}. Yellow points in this figure indicate galaxies with negative \(A_{\text{UV}}\) and, consequently, \(A_{\text{V}}\) values. For our analyses, we treat these negative values as zero - implying that these galaxies are `dust free'. We will further elaborate on the rationale and implications of this decision in the subsequent paragraph. From this figure, we find the correlation coefficient is $\sim 0.7$. This discrepancy, resulting from the application of the dust attenuation relation in the UV luminosity method, underscores the necessity for a refined scaling relation, which will likely bring the correlation coefficient closer to equality, assuming that this dust measurement method is the culprit.

\begin{figure}
	\includegraphics[width=\columnwidth]{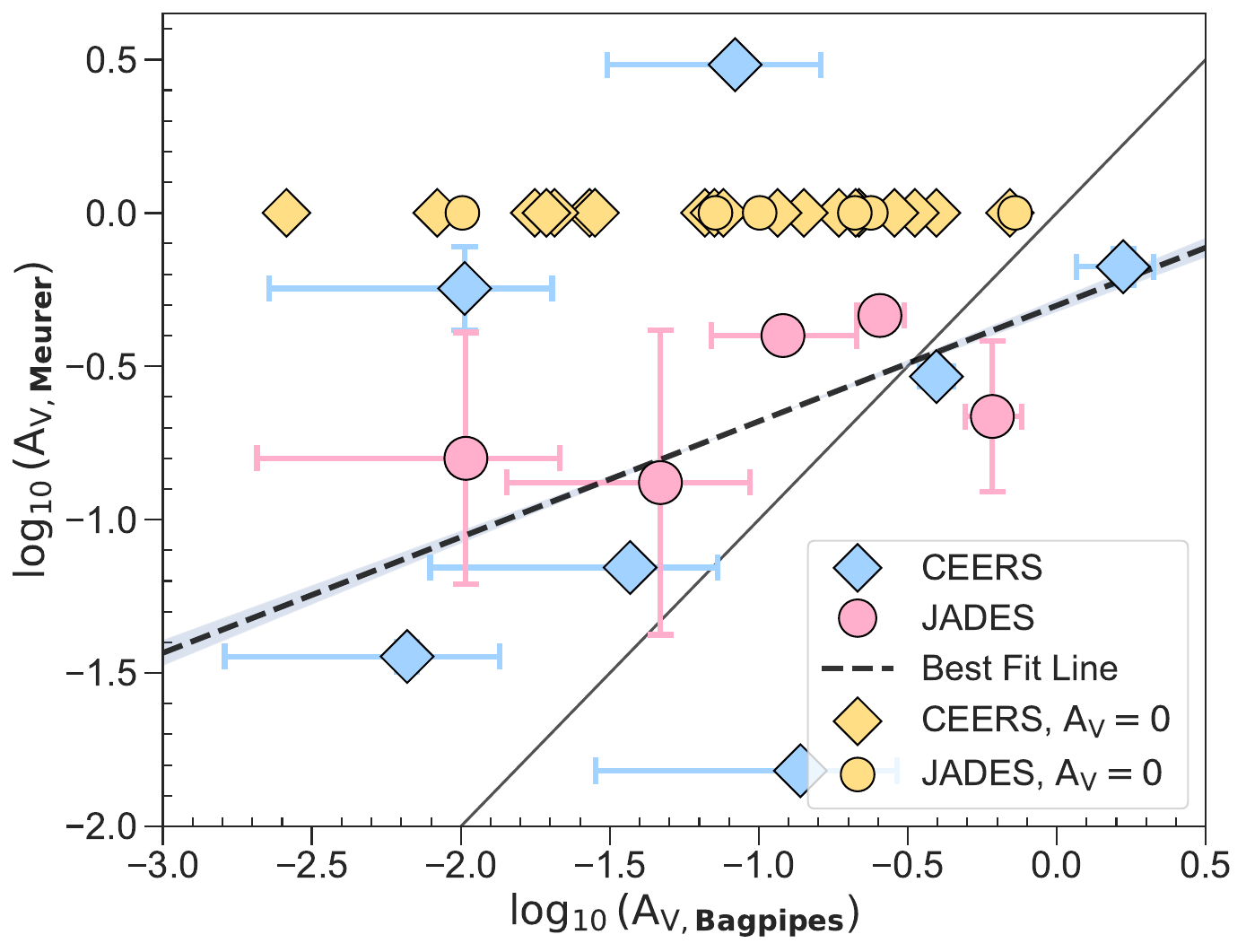}
\caption{Comparison of \(A_{\mathrm{V}}\) values obtained from \protect\cite{1999ApJ...521...64M} and those determined through spectroscopic fitting using \texttt{Bagpipes}. Points marked in yellow represent negative \(A_{\mathrm{V}}\) values as per \protect\cite{1999ApJ...521...64M}, which we reset to zero (dust free). These instances constitute 55\% of the JADES samples and 73\% of the CEERS samples. The best-fit line for data points with positive \( A_{\mathrm{V}} \) values yields a gradient \( m = 0.24 \pm 0.02 \). The 1\(\sigma\) scatter of the residuals, defined as the differences \( x - y \), is measured to be 0.79, indicating high scatter.}

    \label{fig:Av}
\end{figure}

Building upon the above discussion, we detail the values of \(A_{\text{UV}}\) and \(A_{\text{V}}\) utilised in our study to appreciate the scale and implications of our dust corrections. By employing the dust scaling relation from \cite{1999ApJ...521...64M} to determine \(A_{\text{UV}}\) and hence \(A_{\text{V}}\), we find that 55\% of JADES samples and 73\% of CEERS samples exhibit negative \(A_{\text{V}}\) values. These negative values indicate both an absence of dust corrections and issues with calibration; consequently, they are reset to zero.  This is due to the very blue nature of the SEDs of these high-redshift galaxies.  These are bluer than the systems that were used to calibrate the Meurer relation.  The comparison between $\mathrm{A}_\mathrm{V}$ derived from UV $\beta$ slope and \texttt{Bagpipes} is shown in \autoref{fig:Av}. \ Among the galaxies that have positive \(A_{\text{V}}\) values derived from the UV $\beta$ slope, most of the JADES and CEERS sample exhibit \(A_{\text{V}}\) values below 0.5, with a median value of 0.10. Taking into account that 68\% of total samples exhibit negative \(A_{\text{V}}\), this suggests that even for the cases with accurate dust attenuation correction, the magnitude of the necessary correction is typically modest. This conclusion is also shown in \autoref{fig:SFR UV Bagpipes}, where there are no discernible observations showing that the dust-corrected UV SFR deviates from the uncorrected samples.

Our findings underscore the importance of a refined dust scaling relation for high redshift samples. New UV dust scaling relations are being developed, and are needed to make progress on this front. For instance, a comprehensive cosmological hydrodynamical simulation of dust attenuation is presented in \cite{2018MNRAS.473.5363W}. Moreover, a promising technique to recover the dust content of galaxies using machine learning methods is being explored (Fu et al. 2023, in prep). Concurrently, a new empirical relation is also under construction (Austin et al. 2023, in prep).

\subsubsection{Relations of SFR, Mass and Redshift}

In this section we aim to determine the relationship between SFR, masses, and redshifts for our samples. To illustrate our findings, we have created several figures.

\begin{figure}
	\includegraphics[width=\columnwidth]{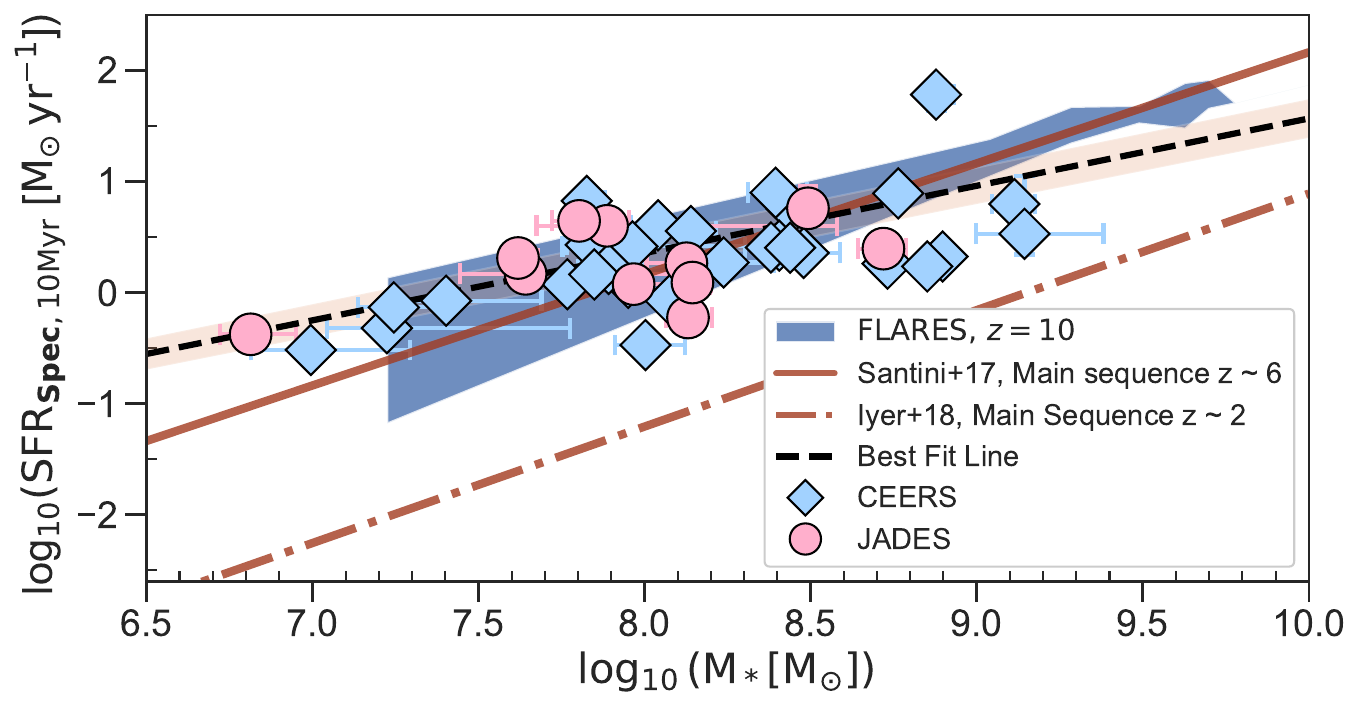}
    \caption{Plot of SFR from \texttt{Bagpipes} with a 10 Myr timescale versus stellar masses. Results from the FLARES simulation with a 10 Myr timescale and main sequence relation at $z \sim 2$ \citep{iyer2018sfr} and $z \sim 6$ \citep{santini2017star} are also shown. The best-fit line is characterized by \( 0.61 \pm 0.01 \). Despite the slightly lower gradient in our results, close agreement with these established studies is observed.}
    \label{fig:sfr mass}
\end{figure}

\autoref{fig:sfr mass} presents the plot of SFR versus stellar masses, derived from \texttt{Bagpipes} using a short 10 Myr SFR timescale. This is compared with the results from the FLARES simulation \citep{wilkins2022first} and the main sequence relations at \( z \sim 2 \) \citep{iyer2018sfr} and \( z \sim 6 \) \citep{santini2017star}. Our findings are in close alignment with these three established studies. We determine a best-fit line, represented by \(y = 0.61x - 4.49\), and find that the 1\(\sigma\) scatter of the residuals is 0.43, indicating low scatter. Although the gradient of our best-fit line is less steep than those found in the aforementioned studies, it should be noted that this discrepancy may be attributable to bias in the selection of our sample.

\begin{figure*}
	\includegraphics[width=\textwidth]{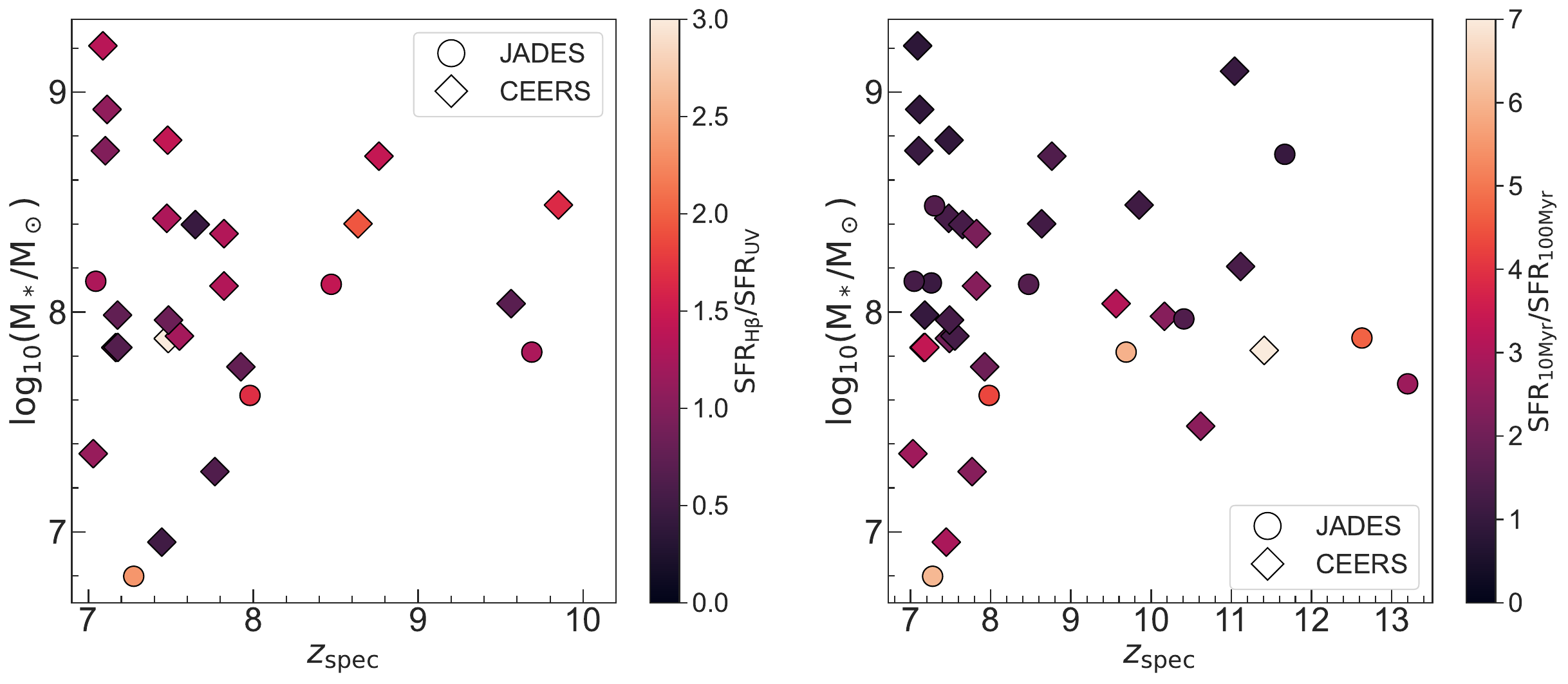}
    \caption{Scatter plots depicting the relationship between stellar masses and the redshift, with color coding representing the SFR ratio (10 Myr / 100 Myr) values. Stellar masses are derived from \texttt{Bagpipes} in both plots. The left plot showcases the SFR calculated using the H$\beta$ line emission and UV luminosity methods, while the right plot displays the SFR as determined by \texttt{Bagpipes} over 10 Myr and 100 Myr timescales. Only the \texttt{Bagpipes}-derived SFR adopts a dust correction factor. The left figure has fewer data points because not all galaxies exhibit an H$\beta$ emission line (26/43). As can be seen from the right plot, galaxies with higher masses tend to have more comparable SFR derived between 10 Myr and 100 Myr timescales.}
    \label{fig:z mass sfr}
\end{figure*}

\begin{figure*}
	\includegraphics[width=\textwidth]{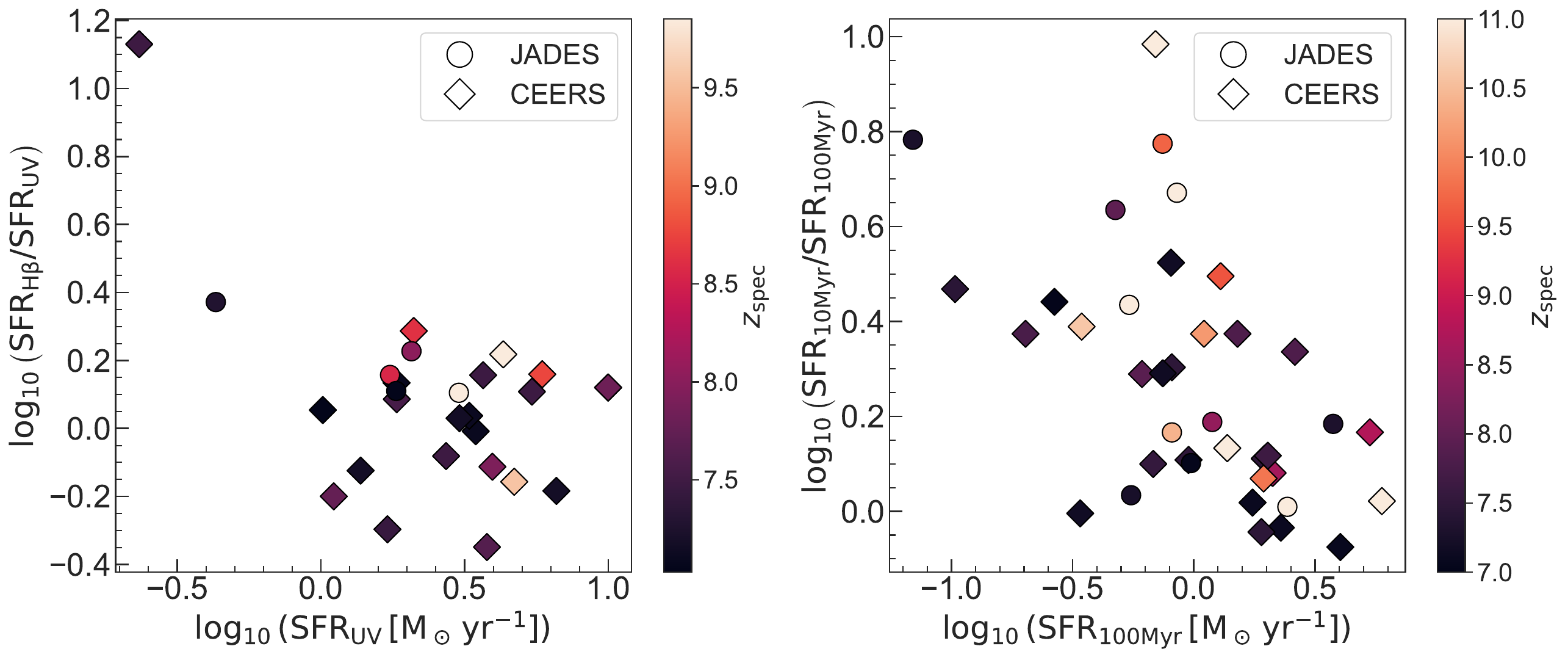}
    \caption{Plots akin to \autoref{fig:z mass sfr}, but with the y-axis representing the SFR ratio and the x-axis displaying the average SFR over a 100 Myr timescale, while the color denotes redshifts. The SFR ratio is more noticeable for galaxies with lower average SFR during the past 100 Myr.}
    \label{fig:sfr sfr z}
\end{figure*}

We present two sets of scatter plots that illustrate the relationships among SFR, stellar masses, and redshifts, as shown in \autoref{fig:z mass sfr} and \autoref{fig:sfr sfr z}. Each set contains two sub-plots: in the left sub-plot, the SFR is calculated using H$\beta$ line emission and UV luminosity, while the stellar mass is derived using \texttt{Bagpipes}. In the right sub-plot, both the SFR and stellar mass are determined via \texttt{Bagpipes}. We subsequently compute the ratio \( \mathrm{SFR_{\mathrm{H}\beta}} / \mathrm{SFR_{\mathrm{UV}}} \) for the left plot, and \( \mathrm{SFR}_{\mathrm{10Myr}} / \mathrm{SFR}_{\mathrm{100Myr}} \) for the right plot, for further analysis. Dust corrections are only considered in the \texttt{Bagpipes} case. 

From the right panel (derived SFR using \texttt{Bagpipes}) of   \autoref{fig:z mass sfr}, it is evident that more massive galaxies generally exhibit comparable SFR values derived from both 10 Myr and 100 Myr timescales, consistent across all redshifts in our samples. This demonstrates the absence of a significant recent burst in SFR for high-redshift galaxies that are more massive \(\log_{10}(\text{M}_* / \text{M}_{\odot}) > 8.6\). However, this observation is not mirrored in the left panel which might be largely attributable to the absence of a dust attenuation correction for the SFR derived from UV and H$\beta$ luminosity. If an accurate dust scaling relation for UV luminosity is developed, then we expect the left result to be similar to the right.

From \autoref{fig:sfr sfr z}, we find that the $\text{SFR}_{10\, \text{Myr}} / \text{SFR}_{100\, \text{Myr}}$
ratio is higher on average for galaxies with a lower SFR as determined by the 100 Myr timescale. The results from this figure's left and right images support this observation. This underscores the recent bursty star-formation patterns, and such bursty star formation histories are particularly pronounced in younger and less massive galaxies, aligning with the findings of \cite{2023arXiv230602470L}.    Furthermore, we do not observe any significant correlations between redshifts and either stellar mass or SFRs for our sample galaxies in the range \( z_\text{spec} = 7 - 13.2 \). This suggests that galaxies within this high-redshift interval may exhibit a diverse range of behaviors.

\subsection{Emission Line Characteristics}
\label{sec: EW}
We investigate the emission line attributes in the four distinct JADES galaxies that prominently display strong H\(\beta\)  \(\lambda4861\), [O~\textsc{iii}] \(\lambda4959\), and [O~\textsc{iii}] \(\lambda5007\) emission lines, using the \texttt{specutils} package \citep{2019ascl.soft02012A}. These lines, within the NIRSpec wavelength range coverage, exhibit the strongest S/N ratio compared to other potential lines. Our choice of these galaxies is informed by two primary factors. Firstly, these JADES galaxies have longer NIRSpec exposure times than the CEERS galaxies, leading to a superior S/N ratio. Secondly, of the 13 JADES galaxies with $z_\text{spec} > 7$,  two systems are without NIRcam images, and 4 (at $z$ = 10.3 - 13.2) are identified as metal-poor galaxies \citep{curtis2022spectroscopic}. Among the remaining 7, only 4 of these galaxies distinctly exhibit the aforementioned three emission lines. The associated spectra for these galaxies are laid out in Appendix \ref{appen:4 JADES Spectrum}, and \autoref{tab:JADES_emission_lines} shows the line flux and equivalent width (EW) of these three lines.

To compare the spectra of these systems with their photometry we attempt to estimate equivalent widths from the photometry. This is a technique to learn about galaxy emission lines without spectra, something which has been done using Spitzer photometry to determine properties of high redshift galaxies \citep[e.g.,][]{Smit2016ApJ}.  To test this idea using JWST data we compare the sum of the equivalent widths for these three lines as derived spectroscopically with their photometric counterparts. The computation of photometric equivalent widths hinges on the differential broad-band magnitudes, specifically between the bands featuring emission lines and those devoid of them. The aggregate equivalent width inherent within the band harboring emission lines can be mathematically expressed as:
\begin{equation}
\Delta \text{m} = - 2.5 \log \left(1 + \frac{\text{EW}_\text{Sum} (1 + z)}{\text{Bandwidth}}\right),
\end{equation}

\noindent where $\Delta m$ is the magnitude differences between the filter band with emission line and the continuum, 'Bandwidth' represents the width of the band that includes the emission lines, $\text{EW}_\text{Sum}$ represents the cumulative equivalent width of all emission lines within that filter band. A detailed introduction of this equation is in \cite{duncan2023jwst,2016MNRAS.460.3587M}. This formula succinctly captures the incremental contribution of the emission line to the overall flux of the band. Among our four JADES galaxies, two display emission in the F444W band, using the F410M band as continuum. The other two show emission in the F410M band, with F356W band serving as the continuum.

\begin{figure}
	\includegraphics[width=\columnwidth]{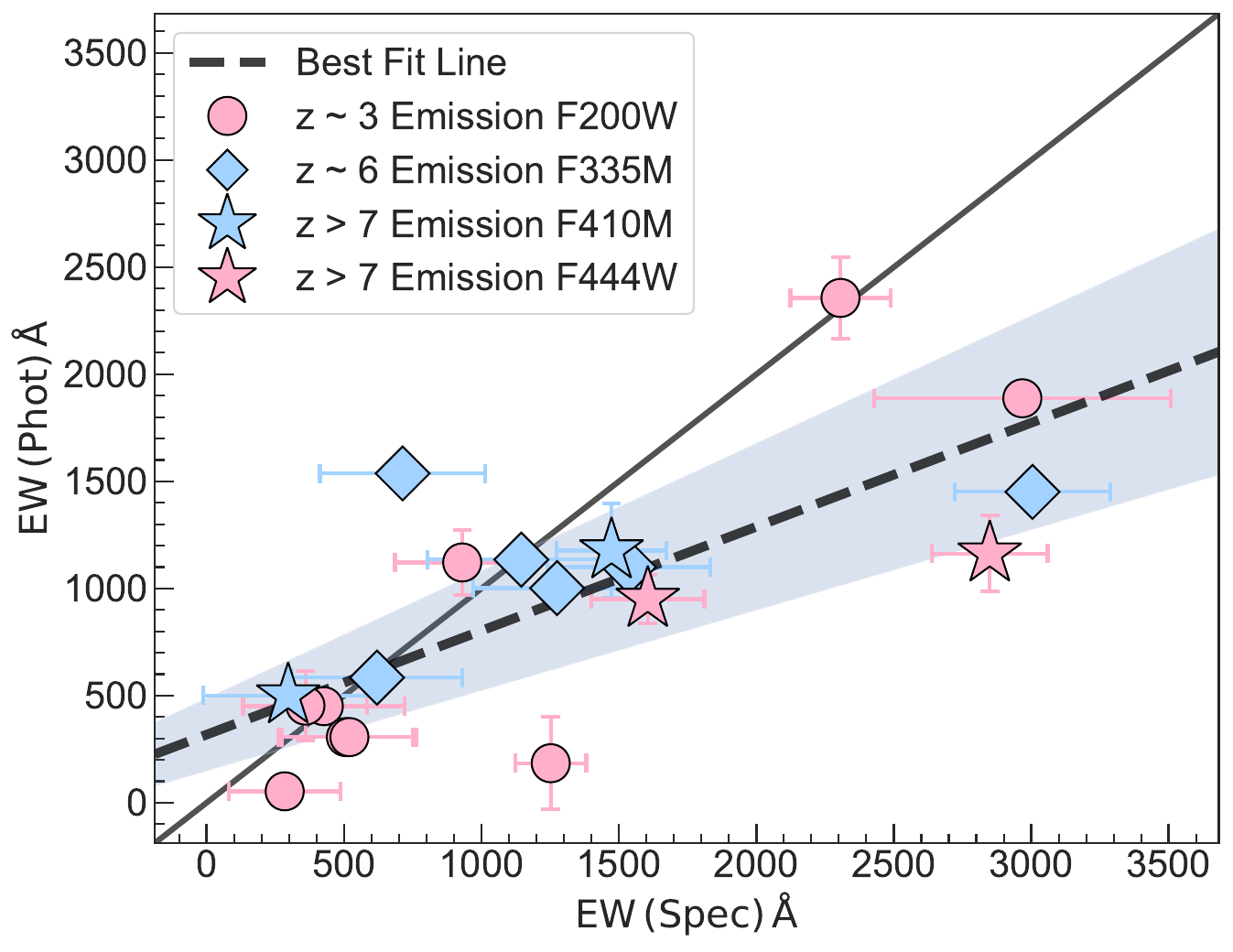}
    \caption{Comparison of the sum of EWs calculated using photometric and spectroscopic methods. The sum represents the combined values of H$\beta$ $\lambda$4861, [O~\textsc{iii}] $\lambda$4959, and [O~\textsc{iii}] $\lambda$5007. The red and blue colors denote the emissions from these three lines in wide and medium filter bands, respectively. Galaxies at different redshifts are labelled with circles, diamonds, and stars for \( z_{\text{spec}} \sim 3, 6, \text{ and } > 7 \)
, respectively.}

    \label{fig:EW all}
\end{figure}

\autoref{fig:EW all} presents a comparative analysis of the sum of the EW of H\(\beta\)  \(\lambda4861\), [O~\textsc{iii}] \(\lambda4959\), and [O~\textsc{iii}] \(\lambda5007\) lines as determined through both photometric and spectroscopic techniques. To ensure a comprehensive study, we incorporate all JADES galaxies at \( z \approx 3 \) that display these three emission lines in the F200W filter. Additionally, six JADES galaxies at \( z \approx 6 \) with these lines detected in the F335M filter are also included.
The gradient of the line of best fit for all samples is $0.49 \pm 0.11$, indicating a moderate agreement between the results obtained from both spectroscopic and photometric approaches. Generally, the photometric method yields sums of EW that are about $30\% \pm 20\%$ lower compared to those derived spectroscopically. We attribute this discrepancy to a potential overestimation of the photometric continuum, leading to diminished EW measurements. While the spectroscopic spectra are uncontaminated, there can be sources of contamination in the photometric data. One possible cause is the assumption that the continuum in the spectrum is flat within the filter band's wavelength range; however, spectra can display various shapes across these wavelengths. In addition, the presence of noise in the spectra can directly influence the size of the continuum, thereby affecting the spectroscopic EW values.

Among the four $z > 7$ JADES galaxies, those with the presence of the three specific emission lines in the F410M medium band (indicated by blue stars in \autoref{fig:EW all}) exhibit more precise photometrically-derived EW values in comparison to galaxies with emission lines in the wide band (F444W). However, this conclusion does not hold as strongly for the $z \approx 3$ samples, which have emissions in the F200W wide band. We believe that the primary underlying factor is still the detection of the continuum. From the spectra of the $z \approx 3$ samples, the continuum is clearly observable and detectable. In contrast, for the four high-redshift samples, the continuum is hardly discernible, as evidenced in Appendix \ref{appen:4 JADES Spectrum}. As a result, when deriving the spectroscopic EW, the continuum introduces uncertainty, leading to deviations from its photometric counterparts.  Given the above considerations, some caution should be used when measuring and interpreting EW measurements from broad-band photometry, especially for galaxies with high EW emission lines. 

Finally, we compare our results with \cite{2023arXiv230411181W}, which studies the sum of the EWs of the same emission lines (H$\beta$ and [O~\textsc{iii}]) for galaxies at redshifts between $1.7$ and $6.7$, and find a  good agreement with our samples within this redshift range.

\begin{table*}
\centering 
\caption{NIRSpec Emission Line Measurements for Four JADES Galaxies: Fluxes and equivalent widths (EWs) for H\(\beta\) \(\lambda 4861\), [O~\textsc{iii}] \(\lambda 4959\), and [O~\textsc{iii}] \(\lambda 5007\) are detailed. Intriguingly, for each galaxy, the ratio of line fluxes does not align with the ratio of their corresponding EWs. This discrepancy may arise from the continuum. The continuum surrounding these emission lines for the four galaxies is scarcely detectable, hence influencing the derived values.}
\label{tab:JADES_emission_lines} 
\begin{tabular}{|c|c|c|c|c|c|c|c|}
\hline
NIRSpec ID & \Large{$z_\text{spec}$} & H\(\beta\) \(\lambda4861\) & [O~\textsc{iii}] \(\lambda4959\) & [O~\textsc{iii}] \(\lambda5007\) & H\(\beta\) & [O~\textsc{iii}] \(\lambda4959\) & [O~\textsc{iii}] \(\lambda5007\) \\
& & Line Flux & Line Flux & Line Flux & EW & EW & EW \\
& & \((10^{-20} \text{ erg s}^{-1} \text{ cm}^{-2})\) & \((10^{-20} \text{ erg s}^{-1} \text{ cm}^{-2})\) & \((10^{-20} \text{ erg s}^{-1} \text{ cm}^{-2})\) & (\AA) & (\AA) & (\AA) \\
\hline
8013 & 8.473 & \(20.59 \pm 2.38\) & \(34.53 \pm 2.83\) & \(93.54 \pm 6.80\) & \(160.3 \pm 72.3\) & \(341.9 \pm 67.4\) & \(1102.6 \pm 66.3\) \\
21842 & 7.98 & \(35.40 \pm 3.13\) & \(64.78 \pm 3.54\) & \(184.81 \pm 3.76\) & \(278.1 \pm 76.1\) & \(620.8 \pm 66.7\) & \(1950.4 \pm 67.4\) \\
20961 & 7.045 & \(46.91 \pm 8.34\) & \(41.97 \pm 6.57\) & \(105.04 \pm 5.40\) & \(18.3 \pm 133.1\)& \(12.2 \pm 92.5\) & \(315.8 \pm 83.5\) \\
10013682 & 7.275 & \(10.10 \pm 2.34\) & \(25.06 \pm 3.11\) & \(61.44 \pm 3.48\) & \(81.4 \pm 57.4\) & \(351.8 \pm 71.4\) & \(1039.8 \pm 70.3\) \\
\hline
\end{tabular}
\end{table*}

\subsection{Morphological and Photometric Size Effects from Line Emission}
\label{sec: morphology}

In our study of the line-emitting sample, we note that the photometric fluxes in line-emitting bands are sometimes stronger than neighboring bands. This brightness can likely be attributed to line emission, as discussed in the previous section. Our primary inquiry in this section is to discern the impact of this line emission on the morphological attributes of galaxies. This is achieved by subtracting and subsequently analyzing the residuals from bands that exhibit line emissions in contrast to those that do not.

A particular focus of our examination are the emission lines \(\text{H}\beta\, \lambda4861\), \([\text{O} \textsc{iii}] \,\lambda4959\), and \([\text{O} \textsc{iii}]\, \lambda5007\), evident in four high-redshift JADES galaxies as discussed in Section 3.3. Of these galaxies, two display the lines in the F444W band (NIRSpec ID: 8013, 21842), while the others do so in the F410M band (NIRSpec ID: 20961, 10013682). To delineate further, the F410M and F356W bands act as the continuum for these sets respectively.  We use these as the continuum as they are the bands closest to those with emission lines, without themselves having emission lines present. Thus, our dataset encompasses two galaxy sets, each offering data from a pair of filter bands – one with emission lines present and its counterpart containing only the continuum.  These can be subtracted from each other to show the location of the line emission spatially.

Our methodology of subtraction is very similar to that used in \citet[][]{Hatch2013}, whereby essentially the line emission structure is found by subtracting a normalised image which contains no lines from the image in the filter where line emission exists.  The idea is that the residuals show the distribution of the gas which produces the line emission.  To do this we carry out a  background subtraction for each image. We do this by masking each galaxy and other galaxies in each image, we then derive the median value for the background level, which is then subtracted from each image. This is followed by the normalization of every galaxy image set, this is a critical step as we have to ensure that all the continuum light is removed from the band with the line emission to reveal that underlying emission. To do this we use an aperture of consistent size across the frames (154 pixels roughly the size of all our galaxies) for each of the galaxies within these images, we compute the total flux within this aperture.  The image with the highest flux summation is used for normalization, from which the normalization constants for other images are determined. The latter is accomplished by dividing the flux summation by their individual flux sums. These constants are then multiplied with the background-subtracted images, resulting in images that are both normalized and devoid of background.

We used this procedure on individual galaxies, however, when this was carried out no single galaxy was found to show line emission that could be detected.  Therefore, we concluded that stacking of these images was potentially a way to retrieve a signal.  To do this for every galaxy set, a weighted stack of these images – both emission and continuum – is created. This involves calculating the standard deviation of the background noise for each image and subsequently assigning weights to each, based on the inverse of the noise standard deviation. 
The final stacked image is constructed by achieving a weighted flux sum and then dividing this by the total weight (the sum of the weights of all images). This procedure is executed separately for the emission and continuum images of every galaxy set. 

\begin{figure}
	\includegraphics[width=\columnwidth]{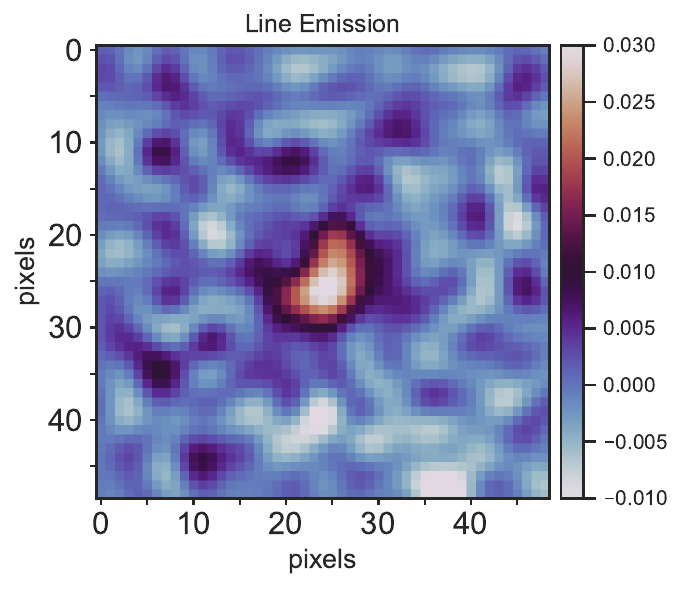}
    \caption{Line emission image obtained by subtracting the stacked continuum-only images from the stacked emission images for the subset of galaxies exhibiting emission lines in the F410M band (NIRSpec ID: 20961, 10013682). A pronounced line emission detection, registering $11.08\sigma$ above the background, is clearly visible, with a possible distinct shape.}

    \label{fig:gas emission}
\end{figure}

To ensure the consistency of the PSF with the F444W band, we employ a two-step process involving the convolution of emission and continuum images with their respective PSF kernels. The PSF models for our bands are generated using \texttt{WebbPSF} \citep{2012SPIE.8442E..3DP, 2014SPIE.9143E..3XP}. The kernels for this convolution are derived using \texttt{pypher} \citep{2016A&A...596A..63B}. These kernels are designed such that when convolved with the PSFs of their specific bands (either emission or continuum), the resultant PSFs are then such that they match that of the F444W band. Due to the emission and continuum residing in different bands, two distinct kernels were crafted and applied for the convolution. After this, the continuum images are subtracted from the emission ones, effectively revealing the location of the material producing the line emission. This assumes that the underlying continuum light in the emission line band is similarly distributed at similar wavelengths.  We test this with measuring the flux below and find a good agreement, revealing that we are indeed retrieving the line emission.  Notably, this emission is accentuated in the galaxy set associated with the F410M band as the emission band, as depicted in \autoref{fig:gas emission}. To quantify the flux of the line emission, eight equal-area apertures are positioned around the emission domain, and the flux sum within these is computed. Through the standard deviation of these sums, we deduce that the core line emission flux sum is elevated at $\sim 11\sigma$ above the background threshold. 

We use these normalization constants to scale the photometric fluxes we measure. Upon analyzing the photometric line flux of this region, as revealed in this image, we obtain a flux measurement of \((203.4 \pm 36) \times 10^{-20} \, \text{erg/s/cm}^2\). This closely aligns with the direct line flux measurements (the sum of the lines in the galaxies stacked), which is found to be \((247.01 \pm 12.86) \times 10^{-20} \, \text{erg/s/cm}^2\), as reported by the JADES team for the same lines in the same galaxies \citep{bunker2023jades}. This is a strong indication that we are indeed seeing the spatial extent of the line emission for these systems, and not as a result of a colour gradient or stellar continuum excess at the emission line band wavelength.

Furthermore, to measure the structure of this line emitting gas we employed the \texttt{GALFIT} software \citep{peng2002detailed, Galfit2} for a detailed morphological analysis. The radii and Sérsic indices of the two galaxies (NIRSpec ID: 20961, 10013682) across different filter bands are presented in \autoref{tab:band_radii}. The photometric band with the stacked line emission has a fitted radius of \(0.61 \pm 0.02\) kpc and a Sérsic index of $n = $ \(0.27 \pm 0.09\). These values align with the average dimensions of the corresponding galaxies in their individual emission bands. Moreover, as emphasized in \autoref{tab:band_radii}, the size of the galaxy gaseous region is slightly larger than stellar contributions, but the errors on these measurements are quite large. Therefore, we can only conclude with this information that the sizes of the emission line regions are statistically similar to the continuum size.   However, the Sérsic index for the line emission image is much lower than for the galaxy continuum images that go into the stack, showing that it is perhaps less concentrated (diffuse) than the stellar light itself.

\begin{figure*} 
    \centering
    \includegraphics[width=\textwidth]{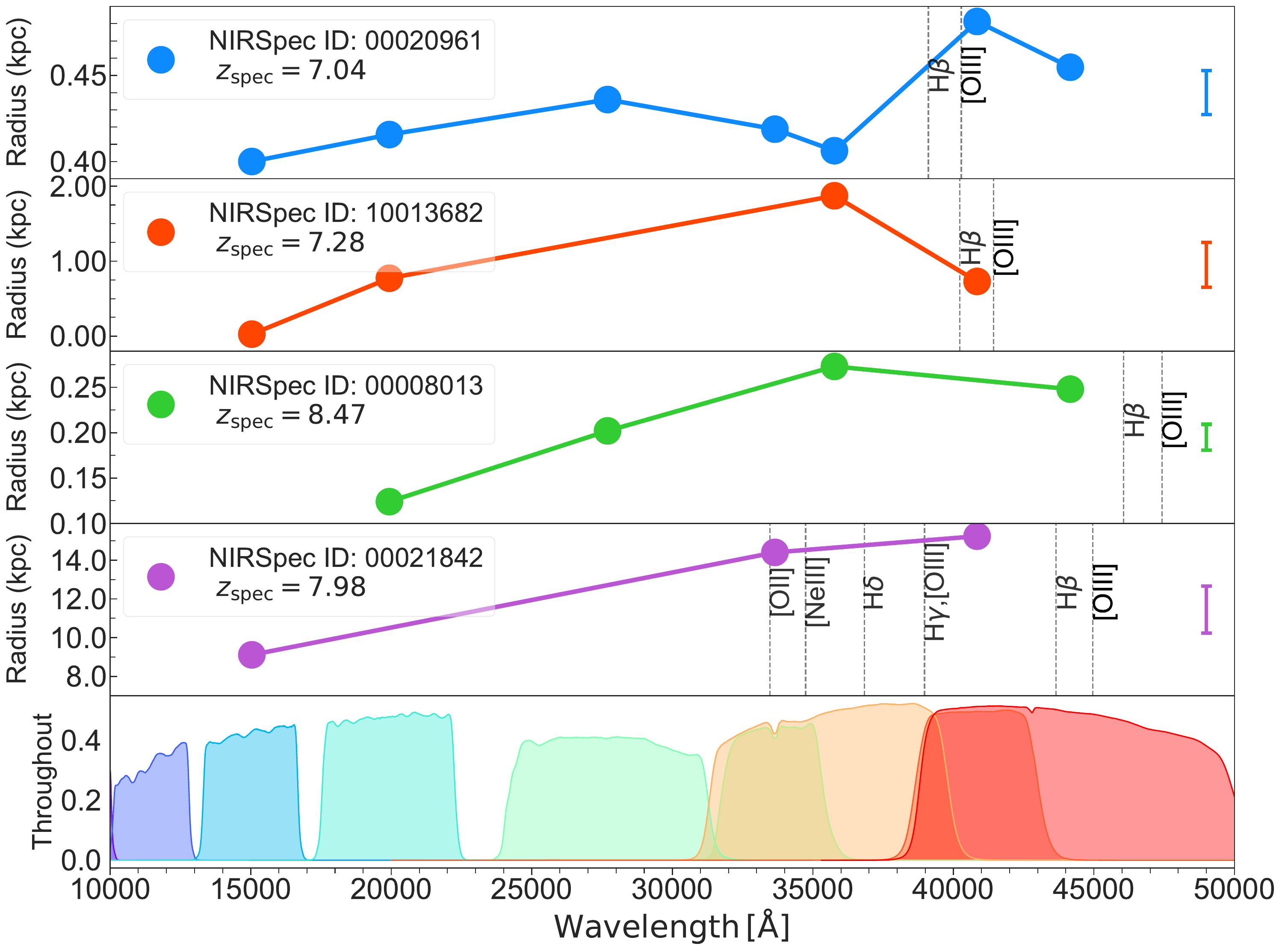}
    \caption{Size comparisons of four JADES galaxies with prominent emission lines. To the right of each individual galaxy plot, the average representative radius error for each galaxy is displayed, while each point on the plots indicates the radius that minimises the $\chi^2_\text{reduced}$ value. We discard any data exhibiting notably high uncertainties or large $\chi^2_\text{reduced}$ values. Typically, the band with the emission line shows a larger radius compared to other filter bands. This implies an extended gas emission region around these galaxies that extends beyond their star-forming regions. The errors are statistically derived from \texttt{GALFIT} and do not necessarily represent physical uncertainties, and are lower limits.}
    \label{fig: JADES 4 Sizes}
\end{figure*}

Lastly, we measure the sizes of the four JADES galaxies with emission lines that overlap in wavelength with the NIRCam filters using \texttt{GALFIT}. After visually inspecting the sizes in these bands, we discard any data exhibiting notably high uncertainties or large $\chi^2_\text{reduced}$ values. The final results are found in \autoref{fig: JADES 4 Sizes}. Notably, we identified a consistent pattern, mirroring findings from the stacked data: bands exhibiting line emission consistently display a slightly larger size relative to those of the continuum bands, with the exception of NIRSpec ID: 10013683.  It is not clear why in that particular case the sizes are not as large.  We do note that in this galaxy, however, we find the weakest emission lines amongst these four systems, which may be the reason. 

\begin{table*}
    \centering
    \caption{Morphological parameters for two JADES galaxies and the attributes of their stacked images are detailed. The stacked residual is calculated by subtracting the Stacked Continuum from the Stacked Emission, highlighting the contribution from gas emission. The uncertainties associated with the radius and Sérsic Index derived from \texttt{GALFIT} are purely statistical, and do not represent physical errors.}
    \begin{tabular}{cccc}
        \hline
        Galaxy & Band & Radius (Kpc) Error & Sérsic Index\\
        \hline
        20961 & Emission & \(0.48 \pm 0.01\) & \(0.59 \pm 0.10\) \\
        20961 & Continuum & \(0.41 \pm 0.01\) & \(0.31 \pm 0.11\) \\
        10013682 & Emission & \(0.73 \pm 0.64\) & \(0.05 \pm 0.29\) \\
        10013682 & Continuum & \(1.87 \pm 0.25\) & \(1.03 \pm 0.50\) \\
        Stacked Emission & Emission & \(0.66 \pm 1.14\) & \(0.03 \pm 0.2\) \\
        Stacked Continuum & Continuum & \(0.49 \pm 1.87\) & \(0.05 \pm 0.72\) \\
        Stacked Residual & Line Emission & \(0.61 \pm 0.02\) & \(0.27 \pm 0.09\) \\
        \hline
    \end{tabular}
    \label{tab:band_radii}
\end{table*}

\section{Discussion}

Our results show that photometric quantities are fairly good at representing the properties of galaxies that can be derived through spectroscopy.    This is under the assumption however, that the quantities we derive from spectroscopy are standard 'correct' values. Whilst this is true for the spectroscopic redshift which is very unlikely to be ambiguous or wrong, this is not necessary the case for star formation and stellar mass, which we discuss below.   Even the measurement of line fluxes for SFR values can be incorrect, despite the common lore that these values are better than others.   It is especially not clear if the measurements of stellar mass and star formation rates are better measured spectroscopically than with photometry.  Under the same assumptions about the underlying process for fitting, that is the same code and same star formation history models, we find that galaxy properties are within 60 \% the same between measurements done with the photometry and spectroscopy for $z > 7$ galaxies.  This is often below the typical random uncertainty limits for these quantities from any measurements we can do now.

We have also shown in this paper that our methods for deriving photometric redshifts using the EPOCHS methods \citep[][]{adams2023discovery} reveal a good agreement with spectroscopic redshift measurements. Obtaining reliable photometric samples is crucial for subsequent spectroscopic redshift follow-up. Given that spectroscopic redshifts are resource-intensive and expensive, we cannot anticipate every galaxy to undergo a spectroscopic analysis due to the associated costs. Consequently, the reliance on photometric redshifts remains paramount for studying the broader galaxy population for the foreseeable future. This dependence is underscored by the fact that these photometric redshifts play a fundamental role in our analyses to decipher evolutionary patterns across various fields. This includes datasets like the PEARLS data \citep{2023AJ....165...13W} and the recent public releases from JWST.  Thus tests such as this one are critical for determining the quality of the photometric redshifts as well as determining what fraction of high redshift galaxies at $z > 7$ would even be included in samples of distant galaxies with photometric redshifts.  One caveat to all of this, which we showed in this paper, is that the spectroscopic samples from JADES and CEERS are quite different in their underlying properties and these certainly are not representative of the distant galaxy population.  More full and complete redshift surveys are needed at these redshifts to determine absolutely how well photometric and selection methods work.

Beyond this we are finding that the gas properties, as measured through emission lines, of these earliest galaxies can be measured with the comparison of spectroscopy and photometry.  This involves extracting the equivalent widths of lines that are present within the photometric bands.  This is the method of finding fluxes or equivalent widths by using the excess in a filter over a fit continuum.   We find that this can be done; however, in some instances, the equivalent widths derived from photometry are about $30\% \pm 20\%$ smaller than those measured with spectroscopy. Our conclusion from this is that any measurements made outside of spectroscopy should be carefully done when trying to measure emission line properties from fluxes within filters.

We also show that new approaches towards understanding galaxy structure in line emission at $z > 7$ can be carried out by subtracting filters with emission lines from those without emission lines to view the entire line emitting structure. We carry this out on a limited sample here, showing that the structure of the gas is slightly diffuse within galaxies. This is an indication that this gas is perhaps not as concentrated as the stars, and gives further evidence for an outside-in formation in these galaxies, assuming that the line emission is produced from star formation events, which from line ratios of these galaxies appears to be the case \citep{rinaldi2023midis, sun2023jades}.

\section{Conclusions}
\label{sec:conclusions}

In this paper we investigate galaxies that have spectroscopy taken with NIRSpec with JWST and are confirmed to be at \(z > 7\). Our primary sample is those galaxies that have NIRSpec data taken as part of the JADES GTO and the CEERS ERS data sets. Our primary goal is to use this spectroscopy and imaging to determine how well photometrically derived quantities, using methods we have developed, compare with those based on the more possibly reliable spectroscopic measurements. Our findings include:

I. We find that there is an excellent agreement in the comparison of photometric redshifts to spectroscopic redshifts using the \texttt{EAZY} code. Only two galaxies are classed as outliers within the full sample of 43 galaxies.  We also discuss in this paper which galaxies in the spectroscopic sample would not be selected using normal procedures for finding high-z galaxies depending on their properties.


II. We find a correlation coefficient \( r \sim 0.60 \) between the stellar masses derived both photometrically and spectroscopically, and a similar correlation for the SFR, using exactly the same \texttt{Bagpipes} setup to measure both. The moderate agreement between results obtained from these two methods underscores the accuracy of the photometric method, given the assumption that spectroscopically derived values are correct.

III. By comparing the star formation rate measurements for our galaxies using the H\(\beta\) line and UV luminosity, we find that there is a 'mismatch' in the spectroscopic properties of the galaxies compared to those derived through photometry. In nearly all cases we find a systematically higher star formation rate (range from ratios of 2.4 to 13.5) as derived through the spectroscopic line fluxes than we get from the photometry itself.  This is an indication that the star formation rate is increasing with time, as the H$\beta$ is measuring more recent star formation.

IV. Furthermore, we find that using broad-band filters to measure emission line equivalent widths is possible, but can lead to high uncertainties and possible underestimates by $30\% \pm 20\%$.  Thus, any measurements of line fluxes or equivalent widths using these filter sets should be done with some caution.

V. We also use a new method to find the spatial distribution of the line emission by subtracting NIRCam filter with and without emission lines present.  Using this method we find that there are no detections of line emission in the individual subtracted images of these galaxies.  However, a stacked version of this method with several galaxies finds a significant detection from which we show that the line emission has a spatial distribution similar to the continuum light.

VI. We measure the morphological and structural properties (size and Sérsic indices) of this sample of galaxies as a function of wavelength in the broad-band and medium-band filters.  We find that in three out of four cases the sizes of these galaxies are slightly larger in the bands that contain the emission lines compared to neighboring bands which are emission line free.  This gives some indication that perhaps the line emission is slightly more extended or less concentrated than the older stellar population.  However, when we subtract off the continuum from the bands with emission lines we find that statistically the sizes of the emission region are similar to the size of the continuum light. 

Overall, we have shown in this paper that the use of photometry to measure galaxy properties is a reliable method of measuring photometric redshifts, stellar masses (or mass to light ratios) and star formation rates. There are slight differences with spectral derived properties and these should be taken into account when trying to calibrate an absolute scale for star formation and stellar mass histories of galaxies which have been derived based on photometry.    In the future, it is clear that more general spectroscopy is needed for early galaxies where tests like these can be done over a broader range of intrinsic properties.

\section*{Data Availability}
Some of the data underlying this article is made available by \cite{2023MNRAS.523..327A}, and the DAWN JWST Archive (DJA). The remainder of the data set will be released together with Conselice et al. (2023, in prep). The catalogues of the sample discussed herein may be acquired by contacting the corresponding author. 

\section*{Acknowledgement}
We thank Elizabeth Stanway for suggestions and thoughts on comparing spectral energy distribution models. 
We acknowledge support from the ERC Advanced Investigator Grant EPOCHS (788113), as well as a studentship from STFC. LF acknowledges financial support from Coordenação de Aperfeiçoamento de Pessoal de Nível Superior - Brazil (CAPES) in the form of a PhD studentship. DI acknowledges support by the European Research Council via ERC Consolidator Grant KETJU (no. 818930). CCL acknowledges support from the Royal Society under grant RGF/EA/181016. CT acknowledges funding from the Science and Technology Facilities Council (STFC). This work is based on observations made with the NASA/ESA \textit{Hubble Space Telescope} (HST) and NASA/ESA/CSA \textit{James Webb Space Telescope} (JWST) obtained from the \texttt{Mikulski Archive for Space Telescopes} (\texttt{MAST}) at the \textit{Space Telescope Science Institute} (STScI), which is operated by the Association of Universities for Research in Astronomy, Inc., under NASA contract NAS 5-03127 for JWST, and NAS 5–26555 for HST. Some of the data products presented herein were retrieved from the Dawn JWST Archive (DJA). DJA is an initiative of the Cosmic Dawn Center, which is funded by the Danish National Research Foundation under grant No. 140. This research made use of the following Python libraries: \textsc{Numpy} \citep{harris2020array}; 
\textsc{Scipy} \citep{2020SciPy-NMeth}; 
\textsc{Matplotlib} \citep{Hunter:2007}; 
\textsc{Astropy} \citep{2013A&A...558A..33A, 2018AJ....156..123A, 2022ApJ...935..167A}; 
\texttt{EAZY-PY} \citep{brammer2008eazy};
\textsc{LePhare} \citep{1999MNRAS.310..540A, 2006A&A...457..841I};
\textsc{Bagpipes} \citep{carnall2018inferring}; 
\textsc{mpi4py} \citep{dalcin2021mpi4py}; 
\textsc{specutils} \citep{2019ascl.soft02012A}; 
\textsc{Pickle} \citep{van1995python}.



\bibliographystyle{mnras}
\bibliography{main} 



\appendix
\section{JADES galaxies Spectrum and Sizes} \label{appen:4 JADES Spectrum}

In this appendix, we present the spectra of four selected JADES galaxies referenced in Sections \ref{sec: EW} and \ref{sec: morphology}. Each figure displays both the simulated and observed photometric flux for each band, with all emission lines labeled. Notably, out of these four galaxies, three exhibit solely the H\(\beta\)  \(\lambda4861\), [O~\textsc{iii}] \(\lambda4959\), and [O~\textsc{iii}] \(\lambda5007\) emission lines. Additionally, we observe slit losses predominantly in the F090W band. We attribute this discrepancy primarily to the F090W band lying blueward of the Lyman break, causing a significant flux drop, especially for our sample galaxies in the redshift range \(z > 7\), which makes the band highly susceptible to noise domination.

\begin{figure*}  
    \centering
    \includegraphics[width=\textwidth]{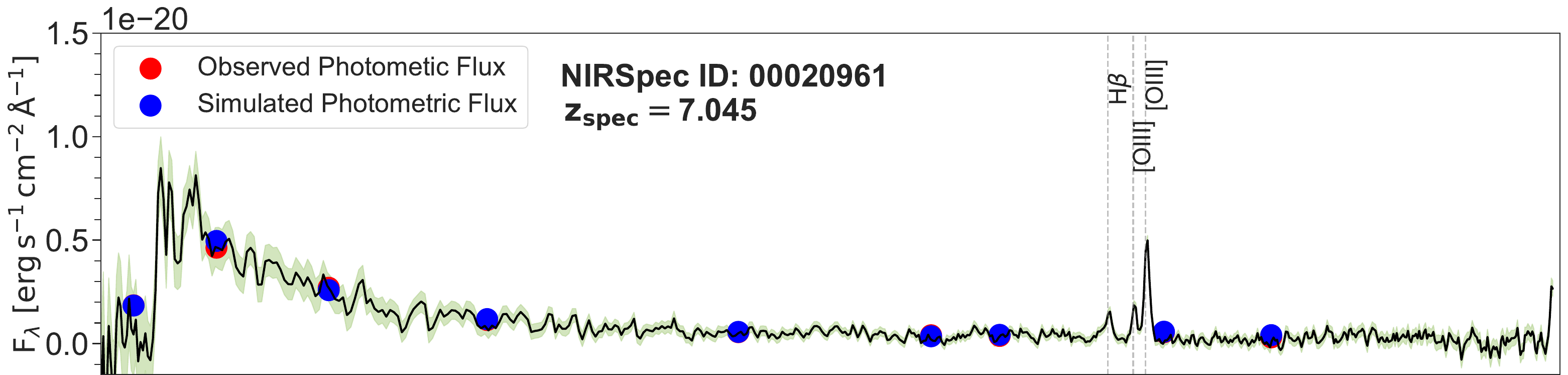}
    \vspace{-\floatsep}  
    
    \includegraphics[width=\textwidth]{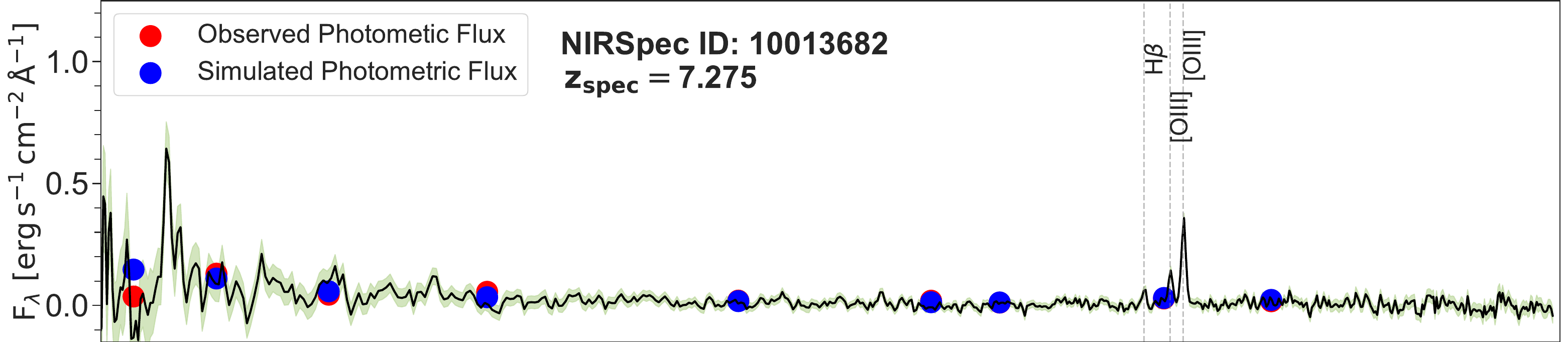}
    \vspace{-\floatsep}  
    
    \includegraphics[width=\textwidth]{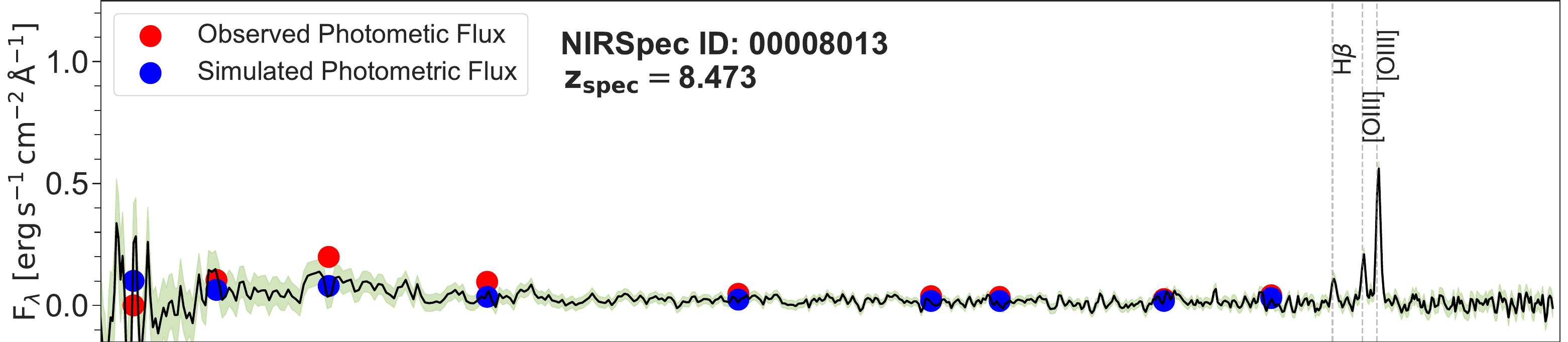}
    \vspace{-\floatsep}  
    
    \includegraphics[width=\textwidth]{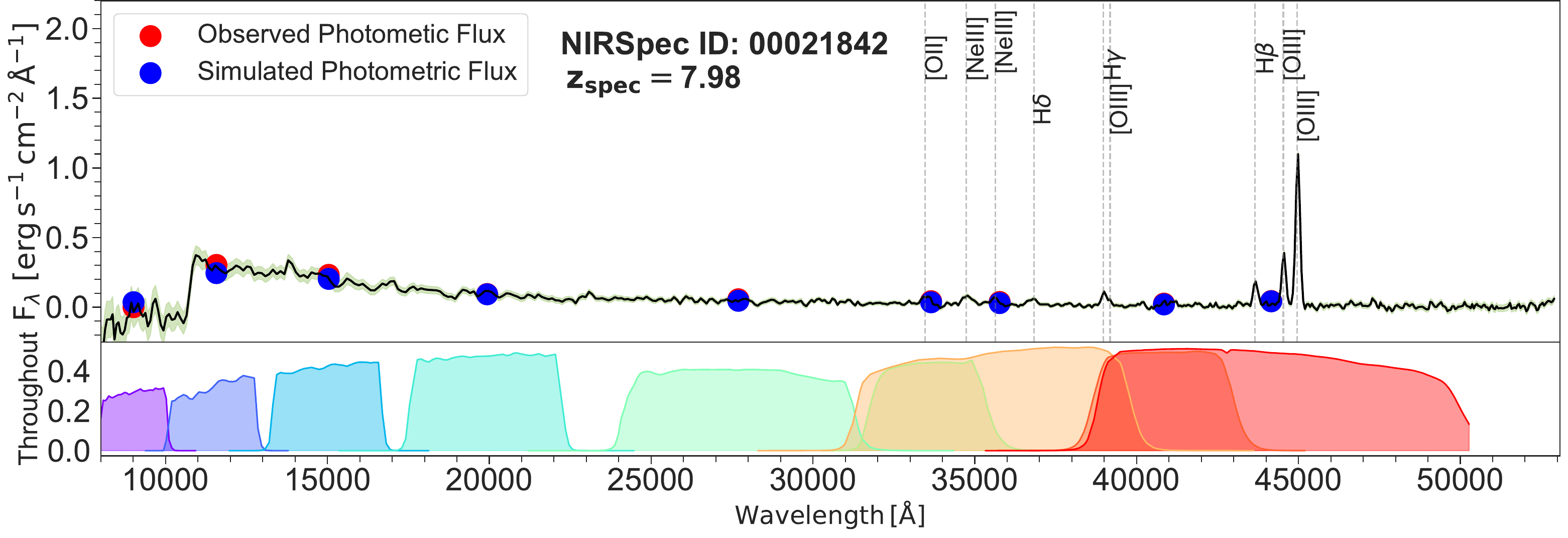}
    \caption{Spectra of four JADES galaxies exhibiting prominent emission lines. Red and blue points denote observed and simulated photometric fluxes, respectively. We find that slit loss occurs only in the F090W band. Additionally, three galaxies exhibit only H\(\beta\) \(\lambda4861\), [O~\textsc{iii}] \(\lambda4959\), and [O~\textsc{iii}] \(\lambda5007\). The relevant line flux and equivalent widths of emission lines are shown in \autoref{tab:JADES_emission_lines}.}
    \label{fig:composite}
\end{figure*}

\bsp	
\label{lastpage}
\end{document}